\documentclass[prb,twocolumn,showpacs,groupedaddress,longbibliography]{revtex4-1}
\usepackage{soul}
\usepackage{epsfig}
\usepackage{subfigure}
\usepackage{graphicx}
\usepackage{amsfonts}
\usepackage{xcolor}
\usepackage[figuresright]{rotating}
\usepackage{amssymb}
\usepackage{amsmath}
\usepackage{psfrag}
\usepackage{esint} 
\usepackage{bm}
\usepackage[colorlinks,linkcolor=blue,anchorcolor=blue,citecolor=blue,urlcolor=blue]{hyperref}
\usepackage{bbm}

\def\be{\begin{equation}} \def\ee{\end{equation}}
\def\bea{\begin{eqnarray}} \def\eea{\end{eqnarray}}

\def\nn{\nonumber}

\def\k{{\bf k}}

\def\d{{\bf d}}
\def\bsigma{{\boldsymbol{\sigma}}}

\makeatletter
\newcommand*{\balancecolsandclearpage}{%
  \close@column@grid
  \clearpage
}
\makeatother

\begin{document}
\title{Berry Phase Enforced Spinor Pairing Order}
\author{Yi Li}
\affiliation{Department of Physics and Astronomy, Johns Hopkins University, Baltimore, Maryland 21218, USA}

\author{Grayson R. Frazier}
\affiliation{Department of Physics and Astronomy, Johns Hopkins University, Baltimore, Maryland 21218, USA}
\date{May 5, 2026}

\begin{abstract}
We introduce a class of topological pairing orders characterized by a half-integer pair monopole charge, leading to Berry phase enforced half-integer partial wave symmetry. This exotic spinor pairing order can emerge from pairing between Fermi surfaces with Chern numbers differing by an odd integer. Using tight-binding models, we demonstrate spinor superconducting orders with monopole charges $\pm 1/2$, featuring a single gap node and nontrivial surface states. Additionally, the superfluid velocity follows a fractionalized Mermin-Ho relation in spatially inhomogeneous pairing orders. The concept extends to spinor density waves and excitons.
\end{abstract}
\maketitle

{\it Introduction.}~--
The discovery of a new pairing order is always accompanied by the establishment of a new paradigm to understand its physical properties and to probe the pairing symmetry. 
It has long been assumed that all pairing orders are completely classified by spherical harmonic symmetries and their lattice counterparts. 
Pioneering examples of unconventional superconductors and superfluids include $p$-wave superfluid $^3$He~\cite{Anderson1961,Balian1963,Leggett1975,Volovik2003}, 
$d$-wave high-$T_c$ cuprates~\cite{Tsuei1994,VanHarlingen1995,Tsuei2000}, and $s_{\pm}$-wave iron pnictides~\cite{Stewart2011,Dai2015}.
Distinct symmetry of pairing gap functions gives rise to characteristic properties in different superconducting states. 
On the other hand, significant progress has been made in the discovery of topological quantum materials characterized by nontrivial geometric phases of single-particle electron bands~\cite{Haldane1988,King-Smith1993,Kane2005a,Fu2007a,Xiao2010} which lead to the discovery of, for example, quantum anomalous Hall insulators~\cite{Yu2010,Chang2013,Haldane2014,Liu2016} and 
Weyl semimetals~\cite{Murakami2007,Wan2011,Xu2011,Yang2011,Burkov2011,Witczak-Krempa2012,Xu2015,Lv2015,Lu2015a,Bradlyn2017,Armitage2018}.

Recently, monopole harmonic superconductivity~\cite{Li2018, Sun2019, Bobrow2022, Frazier2024} has been proposed based on generalization of the single-particle Berry phase to ``\textit{pair} Berry phase" -- the two-particle Berry phase for Cooper pairs in superconductors~\cite{Murakami2003a}. 
This novel topological class of 3D superconductors can possibly exist in, for example, magnetic Weyl semimetals under the proximity effect with an ordinary $s$-wave superconductor. Generally, when pairing occurs between two Fermi surfaces of opposite Chern numbers, the inter-Fermi-surface Cooper pair can nontrivially inherit band topology and acquire a nontrivial ``\textit{pair} Berry phase" in the weak-coupling regime. 
The ``\textit{pair} Berry phase", as a type of topological obstruction, prevents the gap function from being well defined over an entire Fermi surface.
Hence, the corresponding gap function is no longer describable by spherical harmonic functions or their lattice counterparts. Instead, it is characterized by monopole harmonic functions, which are eigenfunctions of angular momentum in the presence of a magnetic monopole~\cite{Dirac1931,Tamm1931,Wu1976,Haldane1983,Li2018}. 
The ``\textit{pair} Berry phase" further enforces gap nodes and determines the total vorticity of gap nodes over a Fermi surface, independent of specific pairing mechanism.
Therefore,  monopole harmonic superconductivity fundamentally differs from previously known unconventional superconductivity.
Furthermore, monopole harmonic superconductivity is only an example of topological many-particle order. In the particle-hole channel, the monopole harmonic charge-density-wave (CDW) order~\cite{Bobrow2020} has been proposed in a model of Weyl semimetal  consisting of nested Fermi surfaces that surround Weyl points of the same chirality.
This novel topological class of CDW states is also characterized by monopole harmonic symmetry and can host emergent Weyl nodes in the gap functions. 

In this article, we first explain the difference between the familiar examples of topological superfluid/superconductors~\cite{Balian1963,Read2000,Volovik2003,Fu2008,Schnyder2008,Chung2009,Lutchyn2010,Qi2011,Alicea2012,Sato2017} and the monopole harmonic superconductors. 
Then, we investigate the spinor pairing as an exotic example of monopole harmonic pairing states. When electrons pair between two topological Fermi surfaces carrying Chern numbers with different even- and oddness, the Cooper pair acquires a half-integer monopole charge. The gap function can exhibit an odd number of nodes over a single Fermi surface, leading to a nontrivial Bogoliubov excitation spectrum. 
We demonstrate the simplest example of an angular momentum $j=1/2$ spinor pairing order using tight-binding models in a cubic lattice, which exhibit a single Bogoliubov-de Gennes (BdG) gap node characterized by nontrivial chirality, providing a nontrivial example of a general result that the Nielsen-Ninomiya theorem~\cite{Nielsen1981a,Nielsen1981b} does not hold in lattice systems when U($1$) symmetry is broken. 
Furthermore, we show zero energy Majorana surface states arising from momentum space phase winding of the low-energy spinor pairing order in the bulk. 
Lastly, we show that the superfluid velocity obeys a fractional generalization of the Mermin-Ho relation~\cite{Mermin1976} in the presence of spatial inhomogeneity of order parameters. 

{\it Monopole harmonic pairing and complex $d$-vectors.}~-- We begin with examples of topological superconductivity where the Fermi surface is topologically trivial, but the superconducting gap function exhibits nontrivial topology.
For example, the spin-polarized $p_x+ip_y$ pairing is fully gapped in 2D. It belongs to class $D$, which breaks both spin-rotation and time-reversal symmetry. 
The gap function $\Delta(\mathbf{k})$ is a complex function exhibiting a phase winding number $\nu = \frac{1}{2\pi}\oint d \k \partial_{\k} \theta (\mathbf{k})$ around the 1D Fermi circle, where $\theta(\k)$ is the U(1) phase of the gap function.
The 3D time-reversal invariant pairing of the $^3$He-B type belongs to class DIII, and the single-particle band structure exhibits two-fold spin degeneracy.
The corresponding gap function is no longer a scalar but is represented by a pairing matrix.
It is proportional to a $2\times 2$ unitary matrix, say, $\Delta(\k)=i\sigma_y \hat \d(\k)\cdot \bsigma$, where the real $d$-vector exhibits a nontrivial texture over the Fermi surface characterized by a nontrivial Pontryagin index,
\bea
\nu=\frac{1}{8\pi}\oiint_{\mathrm{FS}} d k^2 ~ \epsilon_{\mu\nu} \hat \d(\k) \cdot
\partial_{k_\mu} \hat \d(\k) \times \partial_{k_\nu} \hat \d(\k).
\eea
Under an open boundary condition, 2D class D topological superconductors exhibit 1D chiral Majorana modes on the edge, while 3D class DIII time-reversal invariant topological superconductors exhibit helical 2D Majorana surface modes.

Recently, monopole superconductivity~\cite{Li2018} has been proposed as a novel topological class of superconducting pairing order. The central idea is that when Cooper pairing occurs between two Fermi surfaces of different Chern numbers, the resulting Cooper pairs acquire a nontrivial two-particle pair Berry phase, which introduces topological obstruction in the U($1$) phase of the pairing order. 
For example, consider the simplest model of a Weyl semimetal state with a pair of Weyl points located at $\pm\mathbf{K}_0$. 
Upon doping, the Weyl points are enclosed by two separated Fermi surfaces, denoted FS$_{c, \pm}$. 
Near these Fermi surfaces, the low-energy Hamiltonians take the form 
$H_\pm (\mathbf{k} \mp \mathbf{K}_0)=\pm v_F \mathbf{k} \cdot \bsigma -\mu.$ 
The Fermi surfaces FS$_{c, \pm}$ exhibit nontrivial single-particle Berry phases, with single-particle monopole charges being $\pm q=\mp 1/2$.
Alternatively, the corresponding Chern numbers are $\pm C$ with $C=2q$. 
For the pairing between FS$_{c,+}$ and FS$_{c,-}$, the gap function can be represented by a $2 \times 2$ pairing matrix in the spin-$\uparrow,\downarrow$ basis. 
Monopole pairing is an example of \textit{nonunitary} pairing, characterized by complex $\d$-vectors 
$(-d_x+i d_y)/\sqrt 2=u_k^2$, 
$(d_x+i d_y)/\sqrt 2=v_k^2$, $d_z=\sqrt 2 u_k v_k$,
where $u_k=\cos\frac{\theta_{\mathbf{k}}}{2}$ 
and $v_k=\sin\frac{\theta_{\mathbf{k}}}{2} e^{i\phi_{\mathbf{k}}}$. 
The pair Berry connection can be defined in terms of complex $\d$-vectors as
\bea
\mathbf{A}_{pair} (\mathbf{k}) =\hat \d^* \cdot i {\boldsymbol{\nabla}_{\k}} \hat \d
=2q_{pair} \tan\frac{\theta_{\mathbf{k}}}{2}\hat e_{\phi_{\mathbf{k}}}.
\eea
Correspondingly, the Berry curvature can be expressed as $\Omega_i(\mathbf{k})=\epsilon_{ijl}\partial_j A_{pair,l} 
=i\epsilon_{ijl}\partial_j \hat \d^* \cdot
\partial_l \hat \d$. 
The total pair Berry flux through FS$_{c,+}$ is
\begin{equation}
\oiint_{S_+} d\mathbf{k} \cdot \boldsymbol{\nabla}_{k} \times
\mathbf{A}_{pair}(\mathbf{k})=
\oiint_{S_+} d\mathbf{k} \cdot
i\partial_j \hat \d^* \times
\partial_k \hat \d
= 4\pi q_{pair}. \end{equation}
where the pair monopole charges $q_{pair}=2q$.
Hence, the inter-Fermi-surface pairing inherits the Berry phases of electrons residing on different topological Fermi surfaces in a nontrivial way.

As a result, the topological obstruction in the wavefunction of Cooper pairs prevents its phase from being well defined over an entire Fermi surface, leading to generic nodal structures in the pairing gap functions. 
The gap function $\Delta(\mathbf{k})$, when projected onto FS$_{c, \pm}$, exhibits a nodal structure with total vorticity $2q_{pair}$ in momentum space,
which is independent of specific pairing mechanisms~\cite{Li2018}.
When $q_{pair}\neq 0$, $\Delta(\mathbf{k})$ cannot be a regular function throughout the FS$_{c, +}$.
This nodal structure is fundamentally different from that of familiar pairing symmetries based on spherical harmonics $Y_{lm}(\hat{\mathbf{k}})$, which are well defined on the entire Fermi surface and correspond to $q_{pair}=0$ with zero total vorticity.
For example, in the $^3$He-A type $p_x+ip_y$ pairing in 3D, the gap nodes appear at the north and south poles of the Fermi surface as a pair of momentum space vortex and antivortex, respectively.

Away from the Fermi surface, the nodes of $\Delta(\mathbf{k})$ extend into vortex lines in momentum space.
The Weyl points of opposite chiralities are sources and drains for the fundamental vortex lines that intersect the Fermi surface, forming vortices and antivortices in momentum space with a total vorticity of $\pm 2 q_{pair}$. 
Vortex lines that do not connect to Weyl points form closed loops and are nonfundamental - their intersections with the Fermi surfaces create pairs of vortices and antivortices which are not enforced by topology but enrich the symmetry of the pairing order. 
A remarkable feature of this system is that even though these band Weyl points are at high energy near the cutoff scale, far away from the Fermi energy after doping, the non-perturbative nature of topological properties governs the low-energy nodal excitations of $\Delta(\mathbf{k})$ which inherit band topology in a nontrivial way.

{\it Spinor pairing from half-integer pair monopole charges.}~-- So far, we have only discussed monopole pairing with integer-valued monopole charges $q_{pair}$, where the pairing symmetry remains within integer partial-wave channels. 
However, when pairing occurs between two Fermi surfaces with Chern numbers differing by an odd integer, the resulting pair monopole charge $q_{pair}=|C_1-C_2|/2$ becomes a half-integer. 
A notable feature of this pairing is that the order parameter, typically assumed to be bosonic, actually forms a spinor representation of rotation symmetry.

Consider a system consisting of two different types of fermions. The first type is a spin-$1/2$ fermion, with annihilation operators denoted by $c_{\alpha}$, $\alpha=\uparrow,\downarrow$ and mass $m$. This fermion has a 3D Weyl-type spin-orbit coupling as described by $H_c^0$.
The second type of fermion, described by $H_d^0$, is a single-component fermion with annihilation operators denoted by $d$ and mass $M$. It exhibits a simple parabolic dispersion.
\bea
H_c^0&=&\sum_{\mathbf{k}}\sum_{\alpha,\beta=\uparrow,\downarrow} c^\dagger_\alpha(\mathbf{k})
\Big(\frac{\hbar^2k^2}{2m}-\lambda \mathbf{k}
\cdot \boldsymbol{\sigma}_{\alpha \beta}-\mu_c \Big) c_\beta(\mathbf{k}), \nonumber \\
H_d^0&=&\sum_{\mathbf{k}} ~ d^\dagger(\mathbf{k})
\Big(\frac{\hbar^2 k^2}{2M} -\mu_d\Big) d(\mathbf{k}).
\label{band_hamiltonians}
\eea
Here, the system breaks inversion symmetry, resulting in split Fermi surfaces, labeled as FS$_{c,\pm}$, which possess opposite monopole charges, $q_c=\pm 1/2$. 
The Fermi wavevectors, $k_{F;c,\pm}$ and $k_{F;d}$, corresponding to the three Fermi surfaces, satisfy 
$k^2_{F;c,\pm}/2m \pm \lambda k_{F;c,\pm}=\mu_c$ and 
$k^2_{F;d}/2M=\mu_d$.
The synthetic spin-orbit coupling of the $c$ fermion can be possibly realized in ultracold atom systems, which have progressed rapidly from $2$D to $3$D systems, from continuum to lattice models, and from boson to fermion systems~\cite{Lin2011, Li2012,Anderson2012, Wang2012a, Cheuk2012, Song2019, Wang2021}.
Particularly, a Weyl semimetal band was observed in ultracold bosons ($^{87}$Rb) in a $3$D optical lattice, and the spin-orbit coupling can be achieved for ultracold fermions without obstacles in principle~\cite{Wang2021}.

To facilitate pairing between the $c$ and $d$ Fermi surfaces, consider the simplest scenario where the Fermi surface of $d$ fermion, FS$_d$, matches with one of the helical Fermi surfaces of $c$ fermion, say $\mathrm{FS}_{c,+}$. This can be achieved by tuning the chemical potential $\mu_d$ of $d$ fermions such that $k_{F;d}=k_{F;c,+}$. 
In this case, Cooper pairing can occur between FS$_{c, +}$ and FS$_d$, which possess Chern numbers of $-1$ and $0$, respectively.
The mean-field inter-Fermi-surface pairing Hamiltonian takes the form
\bea
H_\Delta=\sum_{\mathbf{k};\alpha=\uparrow,\downarrow} \Delta_\alpha(\mathbf{k})
c^\dagger_\alpha(\mathbf{k}) d^\dagger(-\mathbf{k})
+\Delta_\alpha^*(\mathbf{k})
d (-\mathbf{k})c_\alpha(\mathbf{k}). \ \ \
\label{pairing_hamiltonian}
\eea
Here, the gap function takes a two-component form $\boldsymbol{\Delta}(\mathbf{k})=(\Delta_\uparrow(\mathbf{k}),
\Delta_\downarrow(\mathbf{k}))^\mathrm{T}$. 
Since the kinetic Hamiltonian preserves spin orbit coupled rotational symmetry, the pairing gap function can be expanded in terms of partial wave channels of half-integer angular momentum quantum numbers $j$ and $m_j$ as $\boldsymbol{\Delta}(\mathbf{k})=\sum_{j,m_j} \boldsymbol{\Delta}_{j,m_j}(\mathbf{k})$. 
Due to broken inversion symmetry, each partial wave pairing channel with angular momentum $j$ and $m_j$ is generally a superposition of two orbital angular momentum channels of opposite parity, $l=j-1/2$ and $l+1=j+1/2$: 
$\Delta_{j, m_j;\alpha}(\mathbf{k}) = \Delta_{j, m_j;l,\alpha}(\mathbf{k}) +\Delta_{j, m_j;l+1,\alpha}(\mathbf{k})$ with 
$\Delta_{j, m_j;l,\alpha}(\mathbf{k}) =\Delta_{j, m_j;l}(k)\phi_{j, m_j;l,\alpha}(\hat{\mathbf{k}})$. 
Here, $\Delta_{j, m_j;l}(k)$ is the pairing amplitude and  $\phi_{j, m_j;l,\alpha}(\hat{\mathbf{k}} ) = \langle j, m_j|l, m; \frac{1}{2}, \alpha\rangle Y_{lm} (\hat{\mathbf{k}})\otimes |\alpha\rangle$ is the $\alpha=\uparrow,\downarrow$ component of spin spherical harmonic function~\cite{Edmonds1957}.

To understand the pairing at low energy near Fermi surfaces, we note that  states at $\mathrm{FS}_{c,\pm}$ are respectively described by the helical band eigenstates $|\lambda_\pm(\mathbf{k})\rangle$ for $c$ fermions. They satisfy $\boldsymbol{\sigma} \cdot \hat{\mathbf{k}}|\lambda_\pm(\mathbf{k})\rangle =
\pm |\lambda_\pm(\mathbf{k})\rangle$.
When FS$_d$ matches with FS$_{c,+}$, at low energy, the spin spherical harmonics in the pairing order are projected to monopole harmonics as $P^{(+)} \phi_{j, m_j;l,\alpha}({\hat{\mathbf{k}}} )=
-P^{(+)} \phi_{j, m_j;l+1,\alpha}({\hat{\mathbf{k}}} )=
\frac{1}{\sqrt 2} Y_{q_{pair};j,m_j}({\hat{\mathbf{k}}})$, 
where $q_{pair}=-1/2$, $j=l+1/2$, and $Y_{q_{pair};j,m_j}({\hat{\mathbf{k}}})$ is the monopole harmonic function. It satisfies $J^2 Y_{q_{pair}; j, m_j}=j(j+1) Y_{q_{pair}; j, m_j}$ and $J_z Y_{q_{pair}; j, m_j}= m_j Y_{q_{pair}; j, m_j}$ with $j=|q_{pair}|, |q_{pair}|+1, \cdots$ and $|m_j| \leq j$ (see Supplemental Material (S.M.) S1 for details).
Consequently, the gap function $\boldsymbol{\Delta}(\mathbf{k})$ becomes a function of half charge monopole harmonics, $\Delta^{(+)}(\mathbf{k})= \sum_{j,m_j} \Delta_{j, m_j} Y_{q_{pair}=-1/2;j,m_j}(\hat{\mathbf{k}})$ where $\Delta_{j, m_j}=(\Delta_{j, m_j;l=j-1/2}(k) - \Delta_{j, m_j;l=j+1/2}(k))/\sqrt 2 $. 
The projected pairing Hamiltonian then takes the form
\begin{align}
    H^{(+)}_\Delta(\mathbf{k}) =
    & \sum_{j,m_j} \Delta_{j, m_j} Y_{q_{pair}=-\frac{1}{2};j,m_j}(\hat{\mathbf{k}}) 
    \chi_+^\dagger(\mathbf{k}) d^\dagger(-\mathbf{k}) 
    \nonumber
    \\
    &+ \mathrm{h.c.},
\end{align}
where, $\chi_+^\dagger(\mathbf{k})= \sum_{\alpha=\uparrow,\downarrow} \lambda_{+,\alpha}(\mathbf{k}) c^\dagger_\alpha(\mathbf{k})$ is the creation operator of a single band eigenstate at $\mathbf{k}$ on FS$_{c, +}$.

\begin{figure}[t]
    \centering
    \includegraphics[width=\linewidth]{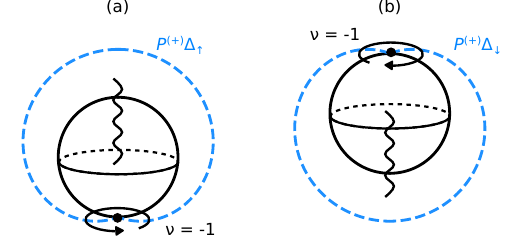}
    \caption{(Color online)
    Spinor pairing gap functions (a)~$P^{(+)}  \Delta_\uparrow \propto Y_{-\frac{1}{2}; \frac{1}{2}, \frac{1}{2}}(\hat{\mathbf{k}})$ and (b)~$P^{(+)}  \Delta_\downarrow \propto Y_{-\frac{1}{2}; \frac{1}{2}, - \frac{1}{2}}(\hat{\mathbf{k}})$ (blue, dashed) overlaid onto Fermi surface $\mathrm{FS}_{c,+}$ 
    (black, solid). 
    They exhibit local chiral $p$-wave symmetry near the gap node under the choice of gauge with Dirac string (wavy line) going through the antipodal point in 
    $\mathrm{FS}_{c,+}$.}
    \label{fig:projected_pairing}
\end{figure}

We emphasize that, regardless of specific interactions in the pairing mechanism, the pairing order always exhibits half-integer charged monopole pairing, following singular spinor representations. 
For instance, consider the simplest case of an attractive contact interaction between these two types of fermions, 
\bea
H_{int}=- g \int d \mathbf{r} c^\dagger_\alpha(\mathbf{r})
 d^\dagger(\mathbf{r}) d(\mathbf{r}) c_\alpha(\mathbf{r}),
 \hspace{1em}\alpha = \uparrow, \downarrow,
 \label{contact_interaction}
\eea
which would only give rise to conventional $s$-wave pairing when Fermi surfaces are topologically trivial. 
In this case, before projection, $\Delta_{\alpha=\uparrow,\downarrow} =-\frac{g}{V}\int d\mathbf{r} \langle G| d(\mathbf{r}) c_\alpha(\mathbf{r}) |G\rangle$ is a constant independent of $\hat{\mathbf{k}}$.
Nevertheless, after the projection, the gap functions $\Delta_{\alpha = \uparrow,\downarrow}$ become
\bea
P^{(+)} \Delta_{\uparrow}&=&  
{\sqrt{2\pi}} \Delta
Y_{-\frac{1}{2};\frac{1}{2},\frac{1}{2}} (\hat{\mathbf{k}})
= \Delta \cos\frac{\theta_{\mathbf{k}}}{2} e^{i\phi_{\mathbf{k}}}, \nn \\
P^{(+)} \Delta_{\downarrow}&=&
{\sqrt{2\pi}} \Delta
Y_{-\frac{1}{2};\frac{1}{2},-\frac{1}{2}} (\hat{\mathbf{k}})
=\Delta \sin\frac{\theta_{\mathbf{k}}}{2} e^{{-}i\phi_{\mathbf{k}}},
\label{projected_pairing.continuum}
\eea
which are time reversal partners with each other.
As shown in Fig.~\ref{fig:projected_pairing}~(a-b), the projected gap function  $P^{(+)}\Delta_{\uparrow}$ ($P^{(+)}\Delta_{\downarrow}$) exhibits a single point node at the south (north) pole on FS$_{c, +}$. 
Locally, we describe the monopole spinor pairing order as familiar chiral $p$-wave by choosing a gauge in which the pairing order is well-defined at the respective nodal point. 
The corresponding single gap node contributes to nonvanishing total vorticity in momentum space,
$\nu=\frac{1}{2\pi} \oint_{C} d\mathbf{k} \cdot \mathbf{v}=2q_{pair}=-1$.
Here,  $\mathbf {v} (\mathbf{k})= \boldsymbol{\nabla}_k \phi(\mathbf{k})-\mathbf{A}_p(\mathbf{k})$ is the gauge-invariant ``velocity" in momentum space. 
Furthermore, we show the energetic stability of the spinor pairing order in S.M.~S3 and S4.

Analogous to the concept of composite fermions~\cite{Jain1989}, this monopole spinor pairing order can be described as a composite order characterized by a local chiral $p$-wave order under a specific gauge choice combined with a nonlocal Dirac string. 
The Dirac string introduces a Berry flux which shifts the angular momentum of the composite pairing order, leading to a spinor representation with gauge invariant half-integer pairing angular momentum. 
Moreover, this monopole spinor pairing order can be extended to describe spinor density waves and spinor excitons in the particle-hole channel.

Next, we study the Bogoliubov quasiparticle excitations. 
As an example, we consider the case of $q_{pair}=-{1}/{2}$, $j=m_j={1}/{2}$, which occurs when inter-Fermi surface pairing between $\mathrm{FS}_{c,+}$ and $\mathrm{FS}_d$ takes the form $\boldsymbol{\Delta} = (\Delta_0, 0)^\mathrm{T}$, with $\Delta_0$ being the pairing amplitude.
In the Nambu basis $\psi(\mathbf{k})=(\chi_+(\mathbf{k}),d^\dagger (-\mathbf{k}))^\mathrm{T}$,
we have
\begin{align}
    \mathcal{H}(\mathbf{k}) 
    = 
    & \nonumber \ 
    \epsilon_{0,k} \tau_0
    +
    \epsilon_k \tau_3 
    +  
    \Delta_0 \sqrt{2 \pi} Y_{-\frac{1}{2}; \frac{1}{2}, \frac{1}{2}}(\hat{\mathbf{k}})\tau_+
    \\
    &+
    \Delta_0 \sqrt{2 \pi} Y^*_{-\frac{1}{2}; \frac{1}{2}, \frac{1}{2}}(\hat{\mathbf{k}})\tau_-,
    \label{BdG_eff.q=1/2,jz=1/2}
\end{align}
where $\tau_\pm = (\tau_x \pm i \tau_y)/2$.
Here,
$\epsilon_{0,k} = (\epsilon_{k; c,+} - \epsilon_{k; d})/2$ and 
$\epsilon_{k} = (\epsilon_{k; c,+} + \epsilon_{k; d})/2$,
with $ \epsilon_{k; c,+} = \hbar^2k^2/(2m) - \lambda k - \mu_c$ and 
$ \epsilon_{k; d} = \hbar^2k^2/(2M) - \mu_d$
being the dispersions of the band eigenstates at the Fermi level.
The Bogoliubov quasiparticle excitations are
$
\gamma^\dagger_1 (\mathbf{k}) = \cos\frac{\beta_k}{2}
\chi^\dagger(\mathbf{k})+\sin\frac{\beta_k}{2} d(-\mathbf{k})$,
$
\gamma^\dagger_2 (\mathbf{k})=\sin\frac{\beta_k}{2} d(\mathbf{k})
- \cos\frac{\beta_k}{2} \chi^\dagger(\mathbf{-k}),
$
where $\tan \beta_k =
\sqrt{2\pi} \Delta_0 Y_{-\frac{1}{2}; \frac{1}{2}, \frac{1}{2}}(\hat{\mathbf{k}})/\epsilon_k$.
The corresponding excitation spectrum is
$E(\mathbf{k})=\epsilon_{0,k} \pm \sqrt{\epsilon_k^2+\Delta_0^2 \cos^2({\theta_{\mathbf{k}}}/{2})}$.
When pairing is instead $\boldsymbol{\Delta} = (0, \Delta_0)^\mathrm{T}$, corresponding to $m_j=-1/2$ channel, we have 
$E(\mathbf{k})= \epsilon_{0,k} \pm \sqrt{\epsilon_k^2+\Delta_0^2 \sin^2({\theta_{\mathbf{k}}}/{2})}$. 
The two pairings exhibit a single BdG node at the south or north
pole, respectively.

\begin{figure}[tb]
    \centering
    \includegraphics[width=\linewidth]{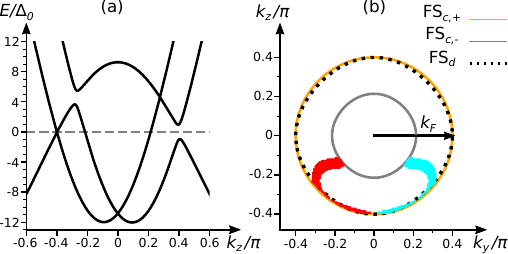}
    \caption{(Color online) 
    (a)~Bulk BdG spectrum of the three-band tight-binding model of spinor superconductor in Eq.~\eqref{tight-binding_model}
    along the $k_z$ axis. 
    It exhibits a single BdG Weyl node at $k_z= - 0.4 \pi$. 
    (b)~zero energy surface 
    arcs (colored) within 2D surface Brillouin zone.
    States localized at open boundary $x=0$ ($x=L_x$) are shown in red (cyan).
    Overlaid are the $k_x=0$ cross sections of bulk Fermi surfaces, FS$_d$ (dashed black line), FS$_{c,+}$ (solid orange line), and FS$_{c,-}$ (solid gray line). 
    Parameters are $t = -1$, $t_d/t= 1$, $t_c/t = 2$, 
    $\mu_d /t= 4.62$,
    $\mu_c/t = 10.38$,
    $\lambda/|t| = 1.20$, $k_F = 0.4 \pi$, and $L_x = 300$.
    The pairing is given by $\boldsymbol{\Delta} = (\Delta_0, 0)^\mathrm{T}$, with $\Delta_0/|t| = 0.15$.
    }\label{fig:tight_binding_model}
\end{figure}
{\it Tight-binding models.}~-- To further demonstrate the exotic spinor pairing and its surface states, we consider a tight-binding Hamiltonian with the kinetic part
$\mathcal{H}^0_{c;t.b.}(\mathbf{k})
    = \sum_i \left( 
    2t_c \cos k_i \mathbbm{1} 
    - \lambda \sin k_i \sigma_i
    \right)-\mu_c \mathbbm{1}$
and $\mathcal{H}^0_{d;t.b.}(\mathbf{k})
    = \sum_i
    2t_d \cos k_i-\mu_d$,
and on-site pairing between $c$ and $d$ fermions. 
It takes the following form,
\begin{align}
    &H_{t.b.}
    = 
    \sum_{\mathbf{n}, \mathbf{n}'}
    \Psi_\mathbf{n}^\dagger
    \left(
    \begin{array}{c|c}
         [\mathcal{H}_c^0]_{\mathbf{n}, \mathbf{n}'} &
         \begin{array}{c}
              \Delta_\uparrow \delta_{\mathbf{n}, \mathbf{n}'}
              \\
              \Delta_\downarrow \delta_{\mathbf{n}, \mathbf{n}'}
         \end{array}
         \\
         \hline
         \begin{array}{cc}
              \Delta_\uparrow^* \delta_{\mathbf{n}, \mathbf{n}'}
              &
              \Delta_\downarrow^* \delta_{\mathbf{n}, \mathbf{n}'}
         \end{array}
         &
         -[\mathcal{H}_d^0]_{\mathbf{n}, \mathbf{n}'}
    \end{array}
    \right)
    \Psi_\mathbf{n'}.
    \label{tight-binding_model}
\end{align}
Here, $\Psi_\mathbf{n} = (c_{\mathbf{n}, \uparrow}, c_{\mathbf{n}, \downarrow}, d^\dagger_\mathbf{n})^\mathrm{T}$, with $c_{\mathbf{n}, \sigma}$ ($d_\mathbf{n}$) being the annihilation operator of $c$ fermion with spin $\sigma$ (spinless $d$ fermion) at site $\mathbf{n}$ in a cubic lattice. 
Nonzero matrix elements of $\mathcal{H}^0_c$ and $\mathcal{H}^0_d$ are
$\left[\mathcal{H}_c^0\right]_{\mathbf{n}, \mathbf{n}} = -\mu_c \mathbbm{1}_{2\times 2}$, 
$\left[\mathcal{H}_c^0\right]_{\mathbf{n}, \mathbf{n} + \boldsymbol{\delta}_i} = t_c \mathbbm{1}_{2\times 2} - \frac{i}{2} \lambda [\sigma_i]_{2\times 2}$, 
$\left[\mathcal{H}_d^0\right]_{\mathbf{n}, \mathbf{n}} =  -\mu_d$, and 
$\left[\mathcal{H}_d^0\right]_{\mathbf{n}, \mathbf{n} + \boldsymbol{\delta}_i} = t_d$.    
Here, $\boldsymbol{\delta}_i$ denotes the direct lattice vector of unit length between nearest neighboring sites along $i = x, y, z$.  
$t_c$ and $t_d$ are spin-independent hopping amplitudes for $c$ and $d$ fermions, respectively, and $\lambda$ is the spin-orbit coupling strength. 
For $\mathcal{H}^0_c$, we take a convention in the tensor product representation that the two-dimensional spin-$\uparrow,\downarrow$ basis is
nested under lattice site basis.
To satisfy the Fermi surface matching condition to pair $\mathrm{FS}_{c,+}$ with $\mathrm{FS}_d$, we take $\mu_d = 2t_d (\cos k_F + 2)$
and 
$\mu_c = 2t_c (\cos k_F + 2) + \lambda |\sin k_F|$,
with $k_F= k_{F;c,+} = k_{F;d}$ being the Fermi wavevector at the overlap of $\mathrm{FS}_{c,+}$ and $\mathrm{FS}_{d}$, as shown in Fig.~\ref{fig:tight_binding_model}~(b).

To demonstrate spinor pairing at 
$m_j = 1/2$ channel, we take $\Delta_\uparrow =\Delta_0$ and $\Delta_\downarrow=0$.
The bulk BdG spectrum of the system is shown in Fig.~\ref{fig:tight_binding_model}~(a), which exhibits a single point node at the south pole of the overlapped Fermi surfaces FS$_{c,+}$ and FS$_{d}$. 
Near the south pole $(0,0, -k_F)$, the low-energy BdG Hamiltonian in the $(\chi_+(\mathbf{k}), d^\dagger(-\mathbf{k}))^\mathrm{T}$ Nambu basis takes the form
$\mathcal{H} (\mathbf{k}) \equiv \mathcal{H}(\tilde{\mathbf{k}} - k_{F} \hat{z})
    \approx
    \left(
    \begin{array}{cc}
        -v_1 \tilde{k}_z
        &  
        v_{\parallel}(\tilde{k}_x + i\tilde{k}_y)
        \\
        v_{\parallel}(\tilde{k}_x -i\tilde{k}_y)
        & v_2 \tilde{k}_z
    \end{array}
    \right)$ with $v_{\parallel}=\Delta_0/(2|\sin k_{F}|)$, $v_1=-2t_c \sin k_{F} + \lambda \cos k_{F}>0$, and $v_2=-2t_d \sin k_{F}>0$.
It exhibits an emergent BdG Weyl node with chirality $-1$.  This is a nontrivial example of a general statement that the Nielsen-Ninomiya theorem does not hold in lattice systems when U($1$) symmetry is broken~\cite{Nielsen1981b}. 

\begin{figure}[tb]
    \centering
    \includegraphics[width=\linewidth]{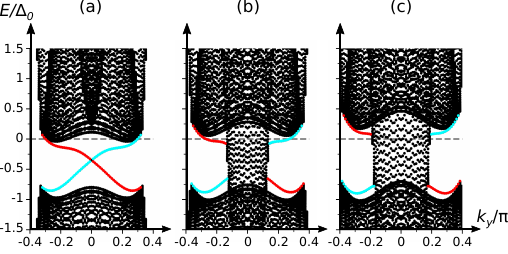}
    \caption{(Color online)
    Energy dispersion of surface modes in Fig. \ref{fig:tight_binding_model}~($b$)
    at (a)~$k_z = -0.22\pi$, (b)~$k_z = -0.16\pi$ where surface modes merge into bulk spectrum (black) for $|k_y|<k_{F;c,-}$, and (c)~$k_z = -0.10\pi$ where only bulk states of unpaired $\mathrm{FS}_{c,-}$ reside at zero energy.
    The color scheme for surface states and parameters are the same as those in Fig.~\ref{fig:tight_binding_model}~($b$).}
    \label{fig:surface_state_dispersion}
\end{figure}

Furthermore, the nontrivial pairing phase winding near the gap node exhibits local chiral $p$-wave symmetry and gives rise to zero energy surface modes localized at opposite sides of the system, as shown in Fig.~\ref{fig:tight_binding_model}~(b). 
These nontrivial surface states originate from the single pairing gap node before merging into the unpaired $\mathrm{FS}_{c, -}$ bulk states,  
as shown in  Fig.~\ref{fig:surface_state_dispersion}~(a-c).
This is a unique characterization of this spinor pairing order and contrasts, for example, a three-dimensional $p_x+ip_y$ superconductor, where the zero energy surface arc connects between the vortex and antivortex at the two poles of the Fermi surface. 
Shown in Fig.~\ref{fig:surface_state_dispersion}~(a-b), there are two distinct surface states localized at $x = 0$ ($L_x$) which have negative (positive) group velocity along $y$ and manifest the local $p_x + ip_y$ symmetry of the pairing order. 
When pairing occurs between $\mathrm{FS}_{c,-}$ and $\mathrm{FS}_{d}$, we have similar results for the spinor pairing with opposite pair monopole charge, as shown in S.M.~S2.

{\it Fractional Mermin-Ho relation.}~-- Lastly, we explore the situation of inhomogeneous spatial distribution of the spinor pairing order. Consider the simplest case where, before projection, the pairing gap function reads $\Delta(\mathbf{r})=|\Delta(\mathbf{r})|e^{i\phi(\mathbf{r})}\eta(\mathbf{r})$, where $\eta(\mathbf{r})$ is a fundamental spin-$1/2$ spinor. 
In momentum space, $\eta$ is projected to the monopole harmonic function $P^{(+)}\eta = \eta_\alpha Y_{-\frac{1}{2};\frac{1}{2},\alpha}(\hat{\mathbf{k}})$. 
Through Hopf map, the spinor gap function is mapped to a unit $3$-vector as
$\hat{\mathbf{n}}(\mathbf{r}) =\eta^\dagger \boldsymbol{\sigma} \eta$.
In $^3$He-$A$ $p$-wave superfluid, the direction of Cooper pairing orbital angular momentum is denoted by the $\mathbf{l}$-vector,
and the curl of the superfluid velocity is determined by the spatial variation of $\mathbf{l}$-vector via
$({\boldsymbol{\nabla} \times 
\mathbf{v}_s}
)_i
=  {(\hbar/2m^*)} \epsilon_{ijk} {\hat{\mathbf{l}} \cdot \partial_j \hat{\mathbf{l}}\times \partial_k \hat{\mathbf{l}}}$ with $m^*$ the mass of the Cooper pair, which is the celebrated Mermin-Ho relation~\cite{Mermin1976}. 
Here, for the spinor pairing order, the corresponding Mermin-Ho relation is fractionalized. The spatial variation of $\hat{\mathbf{n}}$ leads to the following nontrivial circulation of the superfluid velocity (see S.M.~S5 for details)
\bea
(\boldsymbol{\nabla} \times 
\mathbf{v}_s
)_i =
{\frac{\hbar}{4m^*}}
\epsilon_{ijk} \hat{\mathbf{n}} \cdot \partial_j \hat{\mathbf{n}} \times \partial_k
\hat{\mathbf{n}},
\eea
where the $1/2$ factor arises due to the spinor order. 
In a spherical harmonic trap, there will appear a single vortex
on the boundary induced by geometric curvature.

In summary, we have studied a class of topological nodal superconducting states characterized by a nontrivial two-particle pairing Berry phase. 
The low-energy gap function is described by monopole harmonic functions, not only in the integer monopole charge channels but also in more exotic half-integer monopole charge channels. 
The half-integer monopole pairing order follows spinor representation, even though it arises from Cooper pairs which are bosonic. 
The Berry phase enforced spinor pairing order can exist in, for example, ultra-cold atom systems with synthetic spin-orbit coupling, when Cooper pairing occurs between two Fermi surfaces with Chern numbers differing by an odd integer. 
Furthermore, we have demonstrated the $1/2$ monopole charge spinor superconducting order in a lattice. The system exhibits single BdG Weyl node in the bulk spectrum and exotic surface states arising from nontrivial phase winding of spinor pairing order about the gap node. 
Lastly, when the spinor pairing order parameter has a spatial gradient, the superfluid velocity ceases to be irrotational, and a fractional version of the Mermin-Ho relation has been derived. 

{\it Acknowledgments}~--
Y.L. acknowledges the support by the U.S. Department of Energy, Office of Basic Energy Sciences, Division of Materials Sciences and Engineering, Grant No.~DE-SC0019331 and the support of the Alfred P. Sloan Research Fellowships, under which the concept of spinor pairing was developed. Y.L. and G.R.F. acknowledge the support by the NSF CAREER Grant No.~DMR-1848349 for developing the tight-binding model and studying the surface states of the spinor monopole pairing order. 

%

\clearpage
\newpage
\setcounter{equation}{0}
\setcounter{figure}{0}
\setcounter{table}{0}
\setcounter{page}{1}
\makeatletter
\renewcommand{\theequation}{S\arabic{equation}}
\renewcommand{\thefigure}{S\arabic{figure}}
\renewcommand\thesection{S\arabic{section}}
 \renewcommand{\thepage}{S\arabic{page}}

\onecolumngrid

\begin{center}
    \textbf{\large{Supplemental Material for ``Berry Phase Enforced Spinor Pairing Order''}}
    \\
    \vspace{1em}
    Yi Li and Grayson R. Frazier
    \\
    \textit{\small Department of Physics and Astronomy, Johns Hopkins University, Baltimore, Maryland 21218, USA}
\end{center}

\section{Spin spherical harmonics and spinor monopole harmonics}

Consider the mean-field pairing Hamiltonian in Eq.~(5)
in the main text, describing inter-Fermi surface pairing. 
As the system preserves spin-orbit coupled rotational symmetry, the inter-Fermi surface pairing gap function can be expanded into partial wave channels,
$\boldsymbol{\Delta}(\mathbf{k})
=
\sum_{j, m_j}
\boldsymbol{\Delta}_{j, m_j} (\mathbf{k})$. 
Due to broken inversion symmetry, each total angular momentum $j$, $m_j$ channel decomposes into two channels of opposite parity, $l = j- 1/2$ and $l+1 = j+ 1/2$, as
\begin{eqnarray}
    \boldsymbol{\Delta}_{j,m_j} (\mathbf{k}) 
    &=& \boldsymbol{\Delta}_{j, m_j; l} (\mathbf{k}) + \boldsymbol{\Delta}_{j, m_j; l+1} (\mathbf{k}) \nonumber \\
    &=& \Delta_{j, m_j; l}(k)
    \boldsymbol{\phi}_{j,m_j; l} (\hat{\mathbf{k}}) + 
    \Delta_{j, m_j; l+1}(k)
    \boldsymbol{\phi}_{j,m_j; l+1} (\hat{\mathbf{k}}).
\end{eqnarray}
Here 
$\Delta_{j, m_j; l}(k)$ is the pairing amplitude, $\langle j, m_j| l,m; \frac{1}{2}, \alpha \rangle$ with $\alpha = \pm 1/2$ is the Clebsch-Gordan (C-G) coefficient, and $\boldsymbol{\phi}_{j,m_j; l}$ is the spin spherical harmonic~\cite{Edmonds1957SM}, $\boldsymbol{\phi}_{j,m_j; l} (\hat{\mathbf{k}}) 
=\begin{pmatrix}
            \phi_{j_+=l+\frac{1}{2},m_j=m+\frac{1}{2};l,\uparrow}(\hat{\mathbf{k}})\\
            \phi_{j_+=l+\frac{1}{2},m_j=m+\frac{1}{2};l,\downarrow}(\hat{\mathbf{k}})
    \end{pmatrix}
=
    \left(
    \begin{array}{c}
          \langle j, m_j| l,m; 1/2, 1/2\rangle Y_{l, m}(\hat{\mathbf{k}})
         \\
          \langle j, m_j| l,m+1; 1/2, -1/2\rangle Y_{l, m+1}(\hat{\mathbf{k}})
    \end{array}
    \right)$. 
We use shortened notations of the C-G coefficients with spin-$1/2$ particles as
$A_\pm (m)=\left \langle
l\pm \frac{1}{2}, m+\frac{1}{2} | l,m; \frac{1}{2},\frac{1}{2}
\right \rangle
= \pm \sqrt{\frac{1}{2} (1\pm \frac{m+1/2}{l+1/2})}
$ 
and 
$B_\pm (m+1) =\left \langle
l\pm \frac{1}{2}, m+\frac{1}{2} | l,m+1; \frac{1}{2},-\frac{1}{2}
\right \rangle
= \sqrt{\frac{1}{2} (1\mp \frac{m+1/2}{l+1/2})}
$. 
Note that $A_+(m)=B_-(m+1)$ and $A_-(m)=-B_+(m+1)$, which gives the following useful relations,
\begin{equation}
    A_+(m) A_-(m)+B_+(m+1)B_-(m+1)=0, \quad A_+^2(m) + B_+^2(m+1) = A_-^2(m) + B_-^2(m+1)=1.
\end{equation}

The wavefunction of band eigenstate at $\mathrm{FS}_{c,+}$ can be expressed in terms of half-integer monopole harmonics,
\begin{equation}
    \lambda_+ (\hat{\mathbf{k}})
    =
    \sqrt{2\pi}\left(
    \begin{array}{c}
          Y_{\frac{1}{2}; \frac{1}{2}, - \frac{1}{2}}(\hat{\mathbf{k}})
         \\
         -  Y_{\frac{1}{2}; \frac{1}{2}, +\frac{1}{2}}
         (\hat{\mathbf{k}})
    \end{array}
    \right)
    \equiv 
        \left(
    \begin{array}{c}
         u (\hat{\mathbf{k}})
         \\
         v(\hat{\mathbf{k}})
    \end{array}
    \right),
    \label{lambda+.eigstate}
\end{equation}
where, under a choice of gauge, we can have 
$u(\hat{\mathbf{k}})=\cos \frac{\theta_{\mathbf{k}}}{2}$ and 
$v(\hat{\mathbf{k}})=\sin \frac{\theta_{\mathbf{k}}}{2} e^{i \phi_{\mathbf{k}}}$. 
Similarly, the wavefunction of band eigenstate at $\mathrm{FS}_{c, -}$ is given by
\begin{equation}
    \lambda_{-} (\hat{\mathbf{k}})
    =\sqrt{2\pi}
    \left(
    \begin{array}{c}
         - Y_{-\frac{1}{2}; \frac{1}{2}, - \frac{1}{2}}(\hat{\mathbf{k}}) \\ Y_{-\frac{1}{2}; \frac{1}{2}, +\frac{1}{2}} (\hat{\mathbf{k}})
    \end{array}
    \right)
    = 
        \left(
    \begin{array}{c}
         -v^*(\hat{\mathbf{k}}) \\ u^*(\hat{\mathbf{k}})
    \end{array}
    \right),
\end{equation}
where, under the same gauge as that of Eq.~\eqref{lambda+.eigstate}, we have $-v^*(\hat{\mathbf{k}}) =-\sin \frac{\theta_{\mathbf{k}}}{2}
         e^{-i \phi_{\mathbf{k}}}$ and $u^*(\hat{\mathbf{k}})=\cos \frac{\theta_{\mathbf{k}}}{2}$. 
We define the projection operator to the above band eigenstates as $P^{({\pm})} =|\lambda_\pm\rangle \langle\lambda_\pm |$. Since $P^{(+)} + P^{(-)} =1$, the spin spherical harmonic function can be decomposed into its projections to band eigenfunctions as
\begin{eqnarray}
       \boldsymbol{\phi}_{j,m_j; l} (\hat{\mathbf{k}})  
        &=&
        \langle \hat{\mathbf{k}}|\lambda_+\rangle \langle\lambda_+ |\boldsymbol{\phi}_{j; m_j; l}\rangle
        +
        \langle \hat{\mathbf{k}}|\lambda_-\rangle \langle\lambda_-|\boldsymbol{\phi}_{j; m_j; l}\rangle, \quad l=j-1/2, \label{eq:positivehelicitydecomposition}
        \\
        \boldsymbol{\phi}_{j; m_j; l+1} (\hat{\mathbf{k}})   
        &=&
        \langle \hat{\mathbf{k}}|\lambda_+\rangle \langle\lambda_+ |\boldsymbol{\phi}_{j; m_j; l+1}\rangle
        +
        \langle \hat{\mathbf{k}}|\lambda_-\rangle \langle\lambda_-|\boldsymbol{\phi}_{j; m_j; l+1}\rangle, \quad l+1 = j+1/2. 
        \label{eq:negativehelicitydecomposition}
\end{eqnarray}
The projections of spin spherical harmonic function $\boldsymbol{\phi}_{j_+=l+1/2,m_j; l} (\hat{\mathbf{k}})$ to Fermi surfaces $\mathrm{FS}_{c,\pm}$ become 
\begin{eqnarray}
    && P^{(+)}\boldsymbol{\phi}_{j_+=l+1/2,m_j; l} (\hat{\mathbf{k}}) \equiv \langle\lambda_+ |\boldsymbol{\phi}_{j_+=l+1/2; m_j; l}\rangle
    = \begin{pmatrix}
            u^*, v^*
        \end{pmatrix}
        \begin{pmatrix}
            \phi_{j_+=l+\frac{1}{2},m_j=m+\frac{1}{2};l,\uparrow}\\
            \phi_{j_+=l+\frac{1}{2},m_j=m+\frac{1}{2};l,\downarrow}
        \end{pmatrix} \nonumber \\
    &=& \sqrt{\frac{(2l + 1)}{(2l+2)}}
        A_+(m)^2 B_+(0)
        Y_{-\frac{1}{2};  l+\frac{1}{2},  m+\frac{1}{2}}-
        \sqrt{\frac{(2l + 1)}{(2l)}}
        A_+(m) A_-(m) B_-(0)
        Y_{-\frac{1}{2}; l-\frac{1}{2}, m+\frac{1}{2}}  \nonumber \\
        && +
        \sqrt{\frac{(2l + 1)}{(2l+2)}}
        B_+(m+1)^2 B_+(0)
        Y_{-\frac{1}{2};  l+\frac{1}{2},  m+\frac{1}{2}}
        -
        \sqrt{\frac{(2l + 1)}{(2l)}}
        B_+(m+1)
        B_-(m+1)
        B_-(0)
        Y_{-\frac{1}{2}; l-\frac{1}{2}, m+\frac{1}{2}} \nonumber \\
        &=& 
        \sqrt{\frac{2l + 1}{2l+2}} B_+(0)
        Y_{-\frac{1}{2};  l+\frac{1}{2},  m+\frac{1}{2}}\nonumber \\
    &=& \frac{1}{\sqrt{2}}
    Y_{-\frac{1}{2};  l+\frac{1}{2},  m+\frac{1}{2}} . 
    \end{eqnarray}
    \begin{eqnarray}
    && P^{(-)}\boldsymbol{\phi}_{j_+=l+1/2,m_j; l} (\hat{\mathbf{k}})
    \equiv \langle\lambda_-|\boldsymbol{\phi}_{j_+=l+1/2; m_j; l}\rangle = \begin{pmatrix}
            -v, u
        \end{pmatrix}
        \begin{pmatrix}
            \phi_{j_+=l+\frac{1}{2},m_j=m+\frac{1}{2};l,\uparrow}\\
            \phi_{j_+=l+\frac{1}{2},m_j=m+\frac{1}{2};l,\downarrow}
        \end{pmatrix} \nonumber \\
        &=& 
        \sqrt{\frac{(2l + 1)}{(2l+2)}}
        A_+(m)^2
        A_+(0)
        Y_{\frac{1}{2};  l+\frac{1}{2},  m+\frac{1}{2}} -
        \sqrt{\frac{(2l + 1)}{(2l)}}
        A_+(m)
        A_-(m)
        A_-(0)
        Y_{\frac{1}{2}; l-\frac{1}{2}, m+\frac{1}{2}} \nonumber \\
    && +\sqrt{\frac{(2l + 1)}{(2l+2)}}
    B_+(m+1)^2
    A_+(0)
    Y_{\frac{1}{2};  l+\frac{1}{2},  m+\frac{1}{2}}
    -
    \sqrt{\frac{(2l + 1)}{(2l)}}
    B_+(m+1)
    B_-(m+1)
    A_-(0)
        Y_{\frac{1}{2}; l-\frac{1}{2}, m+\frac{1}{2}} \nonumber \\
    &=& \sqrt{\frac{2l + 1}{2l+2}}
    A_+(0)
    Y_{\frac{1}{2};  l+\frac{1}{2},  m+\frac{1}{2}} \nonumber \\
    &=& \frac{1}{\sqrt{2}}
    Y_{\frac{1}{2};  l+\frac{1}{2},  m+\frac{1}{2}}.
\end{eqnarray}
Here, we have employed the formula for the product of two monopole harmonics~\cite{Wu1977SM}, 
\begin{equation}
    \begin{aligned}
        &Y_{q_1;j_1,m_1} Y_{q_2;j_2,m_2} 
        = \sum_{j_3}
        (-1)^{j_1 + j_2 - j_3} 
        \sqrt{\frac{(2j_1 + 1)(2j_2 + 1)}{4\pi(2j_3+1)}}
        \left \langle
    j_1, m_1; j_2, m_2| j_3, m_3
    \right \rangle
    \left \langle
    j_1, q_1; j_2, q_2| j_3, q_3
    \right \rangle
        Y_{q_3; j_3, m_3}.
    \end{aligned}
\end{equation}
Equation \eqref{eq:positivehelicitydecomposition} becomes
\begin{equation}
    \boldsymbol{\phi}_{j_+=l+1/2,m_j; l} (\hat{\mathbf{k}})
    =\begin{pmatrix}
            \phi_{j_+=l+\frac{1}{2},m_j=m+\frac{1}{2};l,\uparrow}(\hat{\mathbf{k}})\\
            \phi_{j_+=l+\frac{1}{2},m_j=m+\frac{1}{2};l,\downarrow}(\hat{\mathbf{k}})
    \end{pmatrix} 
    =
    \frac{1}{\sqrt{2}}
    \left(
    Y_{-\frac{1}{2};  l+\frac{1}{2},  m+\frac{1}{2}}(\hat{\mathbf{k}})
    \begin{pmatrix}
            u(\hat{\mathbf{k}})\\
            v(\hat{\mathbf{k}})
    \end{pmatrix}
    +
    Y_{\frac{1}{2};  l+\frac{1}{2},  m+\frac{1}{2}}(\hat{\mathbf{k}})
    \begin{pmatrix}
            -v^*(\hat{\mathbf{k}})\\
            u^*(\hat{\mathbf{k}})
    \end{pmatrix}
    \right). 
\end{equation}
Similarly, the projections of spin spherical harmonic function $\boldsymbol{\phi}_{j_-=l+1/2,m_j; l+1} (\hat{\mathbf{k}})$ to Fermi surfaces $\mathrm{FS}_{c,\pm}$ become 
\begin{align}
    & P^{(+)}\boldsymbol{\phi}_{j_-=l+1/2,m_j; l+1} (\hat{\mathbf{k}}) \equiv \langle\lambda_+ |\boldsymbol{\phi}_{j_-=l+1/2,m_j; l+1}\rangle
    = \begin{pmatrix}
            u^*, v^*
        \end{pmatrix}
        \begin{pmatrix}
            \phi_{j_-=l+\frac{1}{2},m_j=m+\frac{1}{2};l+1,\uparrow}\\
            \phi_{j_-=l+\frac{1}{2},m_j=m+\frac{1}{2};l+1,\downarrow}
        \end{pmatrix} 
        = -\frac{1}{\sqrt{2}}
    Y_{-\frac{1}{2};  l+\frac{1}{2},  m+\frac{1}{2}} . \\
    & P^{(-)}\boldsymbol{\phi}_{j_-=l+1/2,m_j; l+1} (\hat{\mathbf{k}})
    \equiv \langle\lambda_-|\boldsymbol{\phi}_{j_-=l+1/2; m_j; l+1}\rangle = \begin{pmatrix}
            -v, u
        \end{pmatrix}
        \begin{pmatrix}
            \phi_{j_-=l+\frac{1}{2},m_j=m+\frac{1}{2};l+1,\uparrow}\\
            \phi_{j_-=l+\frac{1}{2},m_j=m+\frac{1}{2};l+1,\downarrow}
        \end{pmatrix} 
        =\frac{1}{\sqrt{2}}
    Y_{\frac{1}{2};  l+\frac{1}{2},  m+\frac{1}{2}}.
\end{align}
Equation \eqref{eq:negativehelicitydecomposition} becomes
\begin{equation}
    \boldsymbol{\phi}_{j_-=l+1/2,m_j; l+1} (\hat{\mathbf{k}})
    =\begin{pmatrix}
            \phi_{j_-=l+\frac{1}{2},m_j=m+\frac{1}{2};l+1,\uparrow}(\hat{\mathbf{k}})\\
            \phi_{j_-=l+\frac{1}{2},m_j=m+\frac{1}{2};l+1,\downarrow}(\hat{\mathbf{k}})
    \end{pmatrix} 
    =
    \frac{1}{\sqrt{2}}
    \left(
    -Y_{-\frac{1}{2};  l+\frac{1}{2},  m+\frac{1}{2}}(\hat{\mathbf{k}})
    \begin{pmatrix}
            u(\hat{\mathbf{k}})\\
            v(\hat{\mathbf{k}})
    \end{pmatrix}
    +
    Y_{\frac{1}{2};  l+\frac{1}{2},  m+\frac{1}{2}}(\hat{\mathbf{k}})
    \begin{pmatrix}
            -v^*(\hat{\mathbf{k}})\\
            u^*(\hat{\mathbf{k}})
    \end{pmatrix}
    \right). \label{eq:positivehelicity_deomposition_final}
\end{equation}
Therefore, when $\mathrm{FS}_d$ matches with $\mathrm{FS}_{c,+}$, the projected pairing gap function takes the form
\begin{eqnarray}
    P^{(+)} \boldsymbol{\Delta}_{j=l+\frac{1}{2},m_j=m+\frac{1}{2}} (\mathbf{k}) 
    &=& P^{(+)} \boldsymbol{\Delta}_{j, m_j; l} (\mathbf{k}) + 
    P^{(+)} \boldsymbol{\Delta}_{j, m_j; l+1} (\mathbf{k}) \nonumber \\
    &=&\Delta_{j, m_j; l}(k)
    P^{(+)} \boldsymbol{\phi}_{j_+=l+1/2,m_j; l} (\hat{\mathbf{k}}) + 
    \Delta_{j, m_j; l+1}(k)
    P^{(+)} \boldsymbol{\phi}_{j_-=l+1/2,m_j; l+1} (\hat{\mathbf{k}}) \nonumber \\
    &=& \frac{1}{\sqrt{2}} (\Delta_{j, m_j; l}(k) - \Delta_{j, m_j; l+1}(k))
    Y_{-\frac{1}{2};  j,  m_j}  (\hat{\mathbf{k}})
    \equiv \Delta^{(+)}_{j, m_j}(k)
    Y_{-\frac{1}{2};  j,  m_j}  (\hat{\mathbf{k}}).
\end{eqnarray}
The projected pairing amplitude in the main text is given by $\Delta_{j, m_j}(k) = \Delta^{(+)}_{j, m_j}(k) = \frac{1}{\sqrt{2}} (\Delta_{j, m_j; l}(k) - \Delta_{j, m_j; l+1}(k))$. 

On the other hand, when $\mathrm{FS}_d$ matches with $\mathrm{FS}_{c,-}$, the projected pairing gap function takes the form
\begin{eqnarray}
    P^{(-)} \boldsymbol{\Delta}_{j=l+\frac{1}{2},m_j=m+\frac{1}{2}} (\mathbf{k}) 
    &=& P^{(-)} \boldsymbol{\Delta}_{j, m_j; l} (\mathbf{k}) + 
    P^{(-)} \boldsymbol{\Delta}_{j, m_j; l+1} (\mathbf{k}) \nonumber \\
    &=&\Delta_{j, m_j; l}(k)
    P^{(-)} \boldsymbol{\phi}_{j_+=l+1/2,m_j; l} (\hat{\mathbf{k}}) + 
    \Delta_{j, m_j; l+1}(k)
    P^{(-)} \boldsymbol{\phi}_{j_-=l+1/2,m_j; l+1} (\hat{\mathbf{k}}) \nonumber \\
    &=& \frac{1}{\sqrt{2}} (\Delta_{j, m_j; l}(k) + \Delta_{j, m_j; l+1}(k))
    Y_{\frac{1}{2};  j,  m_j}  (\hat{\mathbf{k}})
\equiv \Delta^{(-)}_{j, m_j}(k)
    Y_{\frac{1}{2};  j,  m_j}  (\hat{\mathbf{k}}).
\end{eqnarray}
Here, the projected pairing amplitude is $\Delta^{(-)}_{j, m_j}(k)= \frac{1}{\sqrt{2}} (\Delta_{j, m_j; l}(k) + \Delta_{j, m_j; l+1}(k))$.

\section{Tight-binding model of spinor superconductor}
We consider a three-band tight-binding BdG model 
describing a monopole spinor superconductor in a cubic lattice as follows:
\begin{align}
    H &= 
    \sum_{\mathbf{n}, \sigma; \mathbf{n}', \sigma'}
    c_{\mathbf{n}, \sigma}^\dagger
    [\mathcal{H}_{c}^{0}]_{\mathbf{n}, \sigma; \mathbf{n}', \sigma'}
    c_{\mathbf{n}', \sigma'}
    +
    \sum_{\mathbf{n}, \mathbf{n}'}
    d_{\mathbf{n}}^\dagger
    [\mathcal{H}_{c}^{0}]_{\mathbf{n}; \mathbf{n}'}
    d_{\mathbf{n}'}
    +
    \sum_{\mathbf{n}, \sigma}
    \left(\Delta_\sigma c^\dagger_{\mathbf{n}, \sigma} d^\dagger_{\mathbf{n}} +  \mathrm{h.c.}\right).
    \label{tight-binding_model2}
\end{align}
Here, $\mathcal{H}^0_c$ and $\mathcal{H}^0_d$ are the tight-binding band Hamiltonian kernels corresponding to the continuum ones for spinful $c$ fermion and spinless $d$ fermion, respectively. 
$c_{\mathbf{n}, \sigma}$ ($d_\mathbf{n}$) is the annihilation operator of $c$ fermion with spin $\sigma$ (spinless $d$ fermion) at site $\mathbf{n}$.
For matrix elements of $\mathcal{H}_c^0$, we take the following tensor product basis representation, with the two-dimensional spin-$\uparrow,\downarrow$ basis labelled by $\sigma$ nested under lattice site basis indexed by $\mathbf{n}$. 
The nonvanishing elements of $\mathcal{H}^0_c$ are given by
\begin{equation}
    \begin{aligned}
        \left[\mathcal{H}_c^0\right]_{\mathbf{n}, \mathbf{n}} 
        &= 
        -\mu_c \sigma_0, \ \ \
        \left[\mathcal{H}_c^0\right]_{\mathbf{n}, \mathbf{n} + \boldsymbol{\delta}_i} 
        = t_c \sigma_0 - \frac{i}{2} \lambda \sigma_i,
    \end{aligned}
\end{equation}
and the nonvanishing elements of $\mathcal{H}_d^0$ are given by
\begin{equation}
    \begin{aligned}
        \left[\mathcal{H}_d^0\right]_{\mathbf{n}, \mathbf{n}} 
        = 
        -\mu_d, \ \ \
        \left[\mathcal{H}_d^0\right]_{\mathbf{n}, \mathbf{n} + \boldsymbol{\delta}_i} 
        = t_d.
    \end{aligned}
\end{equation}
Above, $\boldsymbol{\delta}_i$ with $i = x, y, z$, are the lattice vectors denoting nearest-neighbour hopping, and spin-independent hopping amplitudes are given by $t_c$ and $t_d$. We take $t_c/2 = t_d = t = -1$. 

As we are considering only inter-Fermi surface pairing, we employ the following reduced Bogoliubov-de Gennes (BdG) Hamiltonian in the $(\mathbf{c}_\uparrow, \mathbf{c}_\downarrow, \mathbf{d}^\dagger)^\mathrm{T}$ Nambu basis,
\begin{equation}
    \mathcal{H}_\mathrm{BdG}
    =
    \left(
    \begin{array}{c|c}
        \mathcal{H}^0_c &
        \begin{array}{c}
             \Delta_\uparrow \\ \Delta_\downarrow
        \end{array}
        \\
        \hline
        \begin{array}{cc}
             \Delta^\dagger_\uparrow & \Delta^\dagger_\downarrow
        \end{array} 
        & - \mathcal{H}^0_d
    \end{array}
    \right),
\end{equation}
with $\mathbf{c}_\uparrow$, $\mathbf{c}_\downarrow$, and $\mathbf{d}^\dagger$ being a shortened notation including site indices.
The Fourier transform of the above tight-binding models is given by
\begin{align}
    \mathcal{H}^0_d(\mathbf{k})
    & = -\mu_d
    + \sum_i
    2t_d \cos k_i,
    \\
    \mathcal{H}^0_c(\mathbf{k})
    & = -\mu_c \sigma_0 
    + \sum_i \left( 
    2t_c \cos k_i \sigma_0
    - \lambda \sin k_i \sigma_i
    \right).
\end{align}
In the continuum limit, the above band Hamiltonians simplify to those in 
Eq.~(4)
of the main text.
In the following tight-binding model, we consider the case when the superconducting pairing is on-site, where $\Delta_{\alpha} = \mathbbm{1} \Delta_\alpha$ with $\alpha = \uparrow, \downarrow$.
Such pairing can occur for a simple contact interaction, as described in Eq.~(7)
of the main text.

\subsection*{Pairing between \texorpdfstring{$\mathrm{FS}_{c,-}$}{FSc-} and \texorpdfstring{$\mathrm{FS}_d$}{FSd}}

To complement the discussion in the main text, we now consider when pairing is instead between $\mathrm{FS}_{c,-}$ and $\mathrm{FS}_{d}$.
This can be achieved by adjusting the chemical potential so that $\mu_d = 2t_d (\cos k_{F;c,-} + 2)$
and 
$\mu_c = 2t_c (\cos k_{F;c,-} + 2) - \lambda |\sin k_{F;c,-}|$, with $k_{F;c,-}$ being the Fermi wavevector of $\mathrm{FS}_{c,-}$.
Because $\mathrm{FS}_{c, -}$ has the opposite Chern number of $\mathrm{FS}_{c,+}$, the pair monopole charge of the Cooper pair is now given by $q_{pair} = 1/2$.
In analogy to Eq.~(9)
of the main text, let us consider the BdG excitations when the inter-Fermi surface pairing is given by $\boldsymbol{\Delta} = (\Delta_0, 0)^\mathrm{T}$, with $\Delta_0$ being the pairing amplitude.
In the continuum limit, the low-energy BdG Hamiltonian in the Nambu basis of $\psi(\mathbf{k})=(\chi_-(\mathbf{k}),d^\dagger (-\mathbf{k}))^\mathrm{T}$
is given by
\begin{equation}
    \mathcal{H}(\mathbf{k})= \epsilon_{0, k} \tau_0 
    + \epsilon_k \tau_3 
    +
    \sqrt{2\pi} \Delta_0 
    Y_{\frac{1}{2}; \frac{1}{2}, \frac{1}{2}}(\hat{\mathbf{k}})
    \tau_+
    +
    \sqrt{2\pi} \Delta_0 
    Y^*_{\frac{1}{2}; \frac{1}{2}, \frac{1}{2}}(\hat{\mathbf{k}})
    \tau_-.
\end{equation}
Above, 
$\epsilon_{0,k} = (\epsilon_{k; c,-} - \epsilon_{k; d})/2$ and 
$\epsilon_{k} = (\epsilon_{k; c,-} + \epsilon_{k; d})/2$,
with $ \epsilon_{k; c,-} = \hbar^2k^2/(2m) + \lambda k - \mu_c$ and 
$ \epsilon_{k; d} = \hbar^2k^2/(2M) - \mu_d$
being the dispersions of the band eigenstates at the Fermi level.
Here, the low-energy pairing order has the symmetry of the monopole harmonic $Y_{\frac{1}{2}; \frac{1}{2}, \frac{1}{2}}(\hat{\mathbf{k}}) = \sin({\theta_{\mathbf{k}}}/{2}) e^{i\phi_{\mathbf{k}}}$, which has a single point node at the north pole.
Near the gap node, the pairing order has local $p_x + ip_y$ symmetry, and the total vorticity is given by $\nu = 2q_{pair} = 1$.

\begin{figure}[bth]
    \centering
    \includegraphics[width=.5\linewidth]{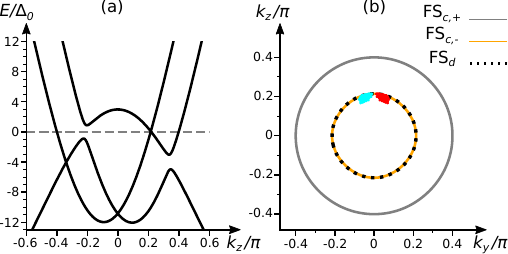}
    \caption{
    (a)~Bulk BdG spectrum of the three-band tight-binding model of spinor superconductor
    along $k_z$ axis. 
    The BdG spectra exhibits a single BdG Weyl node at $k_z= k_{F;c,-} = k_{F;d} = 0.67$. 
    (b)~zero energy surface 
    arcs (colored) within 2D surface Brillouin zone. 
    States localized at $x=0$ ($x=L_x$) are shown in red (cyan).
    Overlaid are the $k_x=0$ cross sections of bulk Fermi surfaces, FS$_d$ (dashed black line), FS$_{c,+}$ (solid gray line), and FS$_{c,-}$ (solid orange line).
    Parameters are the same as that of 
    Fig.~2
    of the main text, but now with $\mu_c/t = 10.38$ and $\mu_d/t = 5.56$, with $k_{F;c,-} = 0.67$ being the wave vector of the paired $\mathrm{FS}_{c,-}$.
    }\label{fig:tight_binding_model.inner_paired}
\end{figure}
In Fig.~\ref{fig:tight_binding_model.inner_paired}~(a), we show the dispersion of the BdG quasiparticle states for the tight-binding model.
Near the south pole, the pairing is fully gapped, whereas as at the north pole,
there is now a single BdG Weyl node.
Due to the Fermi surface matching condition, pairing between $\mathrm{FS}_{c,+}$ and $\mathrm{FS}_d$ is now energetically unfavorable.
The results are analogous to that of Fig.~2
of the main text; only here, due to the Fermi surface matching condition between $\mathrm{FS}_{c,-}$ and $\mathrm{FS}_d$, the BdG node is now at the north pole, at $k_{F;c,-}\hat{z}$.

We study the low-energy excitations of the tight-binding BdG Hamiltonian and demonstrate the single emergent BdG Weyl node.
Expanded near the BdG node at $(0,0, k_{F;c,-})$, the BdG Hamiltonian 
in the Nambu basis of $\psi(\mathbf{k})=(\chi_-(\mathbf{k}),d^\dagger (-\mathbf{k}))$ is given by
\begin{equation}
    \mathcal{H}(\mathbf{k}) \equiv \mathcal{H}(k_{F; c,-} \hat{z}+\tilde{\mathbf{k}}) 
    \approx
    \left(
    \begin{array}{cc}
        -(
        2t_c \sin k_{F; c,-} - \lambda \cos k_{F; c,-}
        ) \tilde{k}_z
        &  
        \frac{\Delta_0}{2|\sin k_{F; c,-}|}
        (\tilde{k}_x + i\tilde{k}_y)
        \\
        \frac{\Delta_0}{2|\sin k_{F; c,-}|}
        (\tilde{k}_x -i\tilde{k}_y)
        & 2t_d \sin (k_{F; c,-}) \tilde{k}_z
    \end{array}
    \right).
\end{equation}
Here, the pairing order has local $p_x + ip_y$ symmetry near the gap node.
In Fig.~\ref{fig:tight_binding_model.inner_paired}~(b), we show the zero energy surface states in the surface Brillouin zone.
Near the projected gap node at $k_{F; c, -} \hat{z}$, the nontrivial pairing phase winding in momentum space gives rise to surface modes localized at opposite surfaces. 
These states originate from the BdG Weyl node but quickly merge into the bulk zero energy states of the unpaired $\mathrm{FS}_{c,+}$. 
The origin of the nontrivial surface states are the same as that discussed in the main text,
with the difference being the location of the emergent BdG Weyl node and the net vorticity.


\section{Energetic stability of spinor pairing order}

In this section, we demonstrate the energetic stability of the spinor superconducting order.
Consider the hardcore interaction between $c$  and $d$ fermions, given by
\begin{equation}
    H_\mathrm{int}
    =
    V_0
    \sum_{i}
    c^\dagger_{i, \alpha}
    c_{i, \alpha}
    d^\dagger_{i}
    d_{i},
    \hspace{1em}\alpha = \uparrow, \downarrow,
\end{equation}
in which $V_0$ is the hardcore interaction amplitude. 
In the following, we consider the interaction projected to the states at the Fermi surface,
\begin{equation}
    H_\mathrm{int}
    =
    \frac{1}{N}
    \sum_{\mathbf{k}, \mathbf{k}'}
    V^{(+)}(\mathbf{k}, \mathbf{k}')
    \chi^\dagger_{\mathbf{k}, +}
    \chi_{\mathbf{k}', +}
    d^\dagger_{-\mathbf{k}}
    d_{-\mathbf{k}'},
\end{equation}
in which $\chi^\dagger_{\mathbf{k}, +}$ is the creation operator for eigenstates of $c$-fermions with momentum $\mathbf{k}$ of the spherical Fermi surface participating in the pairing, $V^{(+)}(\mathbf{k}, \mathbf{k}')$ is the projected interaction, and $N$ is the number of states.
Consider the normalized band eigenstate annihilation operator $\chi_{\mathbf{k}, +} = A(\hat{\mathbf{k}}) c_{\mathbf{k}, \uparrow} + B(\hat{\mathbf{k}}) c_{\mathbf{k}, \downarrow}$.
For example, for the continuum model discussed in Eq.~(4) in the main text, $A(\hat{\mathbf{k}}) = \sqrt{2 \pi} Y_{\frac{1}{2}; \frac{1}{2}, -\frac{1}{2}}(\hat{\mathbf{k}})$ and
$B(\hat{\mathbf{k}}) = -\sqrt{2 \pi} Y_{\frac{1}{2}; \frac{1}{2}, \frac{1}{2}}(\hat{\mathbf{k}})$.
The projected interaction can be written as
\begin{equation}
    V^{(+)}(\mathbf{k}, \mathbf{k}') = V_0 \phi(\mathbf{k}) 
    \phi^*(\mathbf{k}'),
\end{equation}
in which $V_0$ is a constant, and $\phi(\mathbf{k})$ is normalized, $\sum_\mathbf{k} |\phi(\mathbf{k})|^2 = 1$.
For $\alpha = \uparrow$, $\phi(\mathbf{k}) = \frac{A(\hat{\mathbf{k}})}{\sqrt{\int \mathrm{d}\Omega_\mathbf{k} |A(\hat{\mathbf{k}})|^2}}$, and for $\alpha = \downarrow$, $\phi(\mathbf{k}) = \frac{B(\hat{\mathbf{k}})}{\sqrt{\int \mathrm{d}\Omega_\mathbf{k}  |B(\hat{\mathbf{k}})|^2}}$.
For Fermi surfaces with nontrivial Chern number, $\phi(\mathbf{k})$ is a monopole harmonic function.

The mean-field Hamiltonian, after projection to the band eigenbasis, is given by
\begin{align}
    \nonumber
    H_\mathrm{MF} = 
    &\sum_{\mathbf{k}; \mu=\pm} 
    \xi^{(c)}_{\mathbf{k}, \mu}
    \chi^\dagger_{\mathbf{k}, \mu} 
    \chi_{\mathbf{k}, \mu}
    +
    \sum_{\mathbf{k}} 
    \xi^{(d)}_{\mathbf{k}}
    d^\dagger_{\mathbf{k}}
    d_{\mathbf{k}}
    +
    \\
    &+
    \sum_{\mathbf{k}}
    \left(
    \Delta^{(+)}(\mathbf{k})
    \chi^\dagger_{\mathbf{k}, +}
    d^\dagger_{-\mathbf{k}}
    +
    \Delta^{(+)*}(\mathbf{k})
    d_{-\mathbf{k}}
    \chi_{\mathbf{k}, +}
    \right)
    -
    \sum_{\mathbf{k}, \mathbf{k}'}
    V^{(+)}(\mathbf{k}, \mathbf{k}')
    \langle \chi^\dagger_{\mathbf{k}, +}
    d^\dagger_{-\mathbf{k}}
    \rangle
    \langle
    d_{-\mathbf{k}'}
    \chi_{\mathbf{k}', +}
    \rangle.
\end{align}
Here,
$\chi^\dagger_{\mathbf{k}, \mu}$ is the creation operator for band eigenstates of $c$-fermions with momentum $\mathbf{k}$ and energy $\xi_{\mathbf{k}, \mu}^{(c)}$, in which $\mu = \pm 1$.
The projected gap function is given by
\begin{equation}
    \Delta^{(+)}(\mathbf{k})
    =
    \frac{1}{N}
    \sum_{\mathbf{k}'}
    V^{(+)} (\mathbf{k}, \mathbf{k}')
    \langle d_{-\mathbf{k}'}
    \chi_{\mathbf{k}', +} \rangle
    =
    \frac{V_0}{N} \phi(\mathbf{k})
    \sum_{\mathbf{k}'}
    \phi(\mathbf{k}')
    \langle d_{-\mathbf{k}'}
    \chi_{\mathbf{k}', +} \rangle
    \equiv
    \Delta^{(+)}\phi(\mathbf{k}).
\end{equation}
In the last equality, we have defined 
\begin{math}
    \Delta^{(+)}
    =
    \frac{V_0}{N}
    \sum_{\mathbf{k}'}
    \phi(\mathbf{k}')
    \langle d_{-\mathbf{k}'}
    \chi_{\mathbf{k}', +} \rangle,
\end{math}
which is momentum-independent.
The mean-field Hamiltonian can be expressed as 
\begin{equation}
    H_\mathrm{MF}
    =
    \sum_\mathbf{k}
    \Psi_\mathbf{k}^\dagger
    \mathcal{H}_\mathrm{BdG}(\mathbf{k})
    \Psi_\mathbf{k}
    +
    \sum_{\mathbf{k}}
    \xi^{(c)}_{\mathbf{k}, -}
    \chi^\dagger_{\mathbf{k}, -} 
    \chi_{\mathbf{k}, -}
    -
    \frac{N |\Delta^{(+)}|^2}{V_0},
\end{equation}
in which
\begin{equation}
    \mathcal{H}_\mathrm{BdG}(\mathbf{k})
    =
    \left(
    \begin{array}{cc}
         \xi_{\mathbf{k}, +}^{(c)} & \Delta^{(+)}(\mathbf{k})
         \\
         \Delta^{(+)*}(\mathbf{k}) 
         &
         -\xi^{(d)}_{-\mathbf{k}}
    \end{array}
    \right)
    \label{SM_eq:BdG_kernel}
\end{equation}
is the BdG kernel in the $\Psi_\mathbf{k} = (\chi_{\mathbf{k}, +}, d^\dagger_{-\mathbf{k}})^\mathrm{T}$ Nambu basis.
Eigenstates of the BdG kernel in Eq.~\eqref{SM_eq:BdG_kernel} are given by $\Psi_{\mathbf{k},+} = (u_\mathbf{k}, v_\mathbf{k})^\mathrm{T}$
and
$\Psi_{\mathbf{k},-} = (v^*_\mathbf{k}, -u^*_\mathbf{k})^\mathrm{T}$, in which
\begin{align}
    |u_\mathbf{k}|^2 = \frac{1}{2} \left( 1 + \frac{\bar{\xi}_\mathbf{k}}{E_\mathbf{k}} \right),
    \hspace{2em}
    |v_\mathbf{k}|^2 = \frac{1}{2} \left( 1 - \frac{\bar{\xi}_\mathbf{k}}{E_\mathbf{k}} \right),
    \hspace{2em}
    u_\mathbf{k} v^*_\mathbf{k} 
    =
    \frac{\Delta^{(+)*}(\mathbf{k})}{2E_\mathbf{k}}.
    \label{SM_eq:Bogoliubov_transf}
\end{align}
Here, $\bar{\xi} = (\xi^{(c)}_{\mathbf{k}, +} + \xi^{(d)}_{-\mathbf{k}})/2$ and the BdG energies for states $\Psi_{\mathbf{k}, \pm}$ are
\begin{math}
    \delta\xi_\mathbf{k}
    \pm E_{\mathbf{k}},
\end{math}
in which
\begin{math}
    E_\mathbf{k} = \sqrt{\bar{\xi}_\mathbf{k}^2 + |\Delta^{(+)}(\mathbf{k})|^2}
\end{math}
and
\begin{math}
    \delta\xi_\mathbf{k} = (\xi^{(c)}_{\mathbf{k}, +} - \xi_{-\mathbf{k}}^{(d)})/2.
\end{math}
In the following, we consider the case when the two Fermi surfaces are matched, $\delta \xi_\mathbf{k} \approx 0$.

The energy is given by
\begin{align}
    \nonumber
    \langle H_\mathrm{MF} \rangle=
    E_\mathrm{GS} 
    = \nonumber
    & \sum_{\mathbf{k}; \mu = \pm} \xi_{\mathbf{k}, \mu}^{(c)} 
    \langle 
    \chi^\dagger_{\mathbf{k}, \mu} 
    \chi_{\mathbf{k}, \mu}
    \rangle
    +
    \sum_{\mathbf{k}} \xi_{\mathbf{k}}^{(d)} 
    \langle 
    d^\dagger_{\mathbf{k}} 
    d_{\mathbf{k}}
    \rangle
    \\
    &+
    \sum_{\mathbf{k}}
    \left(
    \Delta^{(+)}(\mathbf{k})
    \langle
    \chi^\dagger_{\mathbf{k}, +}
    d^\dagger_{-\mathbf{k}}
    \rangle
    +
    \Delta^{(+)*}(\mathbf{k})
    \langle
    d_{-\mathbf{k}}
    \chi_{\mathbf{k}, +}
    \rangle
    \right)
    -
    \frac{N |\Delta^{(+)}|^2}{V_0}.
\end{align}
Expressed in terms of the coherence factors, we have
\begin{subequations}
\begin{align}
    \langle 
    d^\dagger_{-\mathbf{k}} 
    d_{-\mathbf{k}}
    \rangle
    &=
    |v_\mathbf{k}|^2(1-n_F(\delta\xi_\mathbf{k} + E_\mathbf{k})) 
    + 
    |u_\mathbf{k}|^2
    (1 - n_F(\delta \xi_\mathbf{k} - E_\mathbf{k}))
    \\
    \langle 
    \chi^\dagger_{\mathbf{k}, +} 
    \chi_{\mathbf{k}, +}
    \rangle
    &=
    |u_\mathbf{k}|^2 n_F(\delta\xi_\mathbf{k} + E_\mathbf{k}) 
    + 
    |v_\mathbf{k}|^2
    n_F(\delta \xi_\mathbf{k} - E_\mathbf{k})
    \\
    \langle 
    \chi^\dagger_{\mathbf{k}, -} 
    \chi_{\mathbf{k}, -}
    \rangle
    &=
    n_F(\xi_{\mathbf{k},-}^{(c)})
    \\
    \langle
    d_{-\mathbf{k}}
    \chi_{\mathbf{k}, +}
    \rangle
    &=
    u_\mathbf{k}^* v_\mathbf{k}
    \Big(
    n_F(\delta \xi_\mathbf{k} + E_\mathbf{k}) - n_F(\delta \xi_\mathbf{k} - E_\mathbf{k})
    \Big)
\end{align}
\end{subequations}
in which $n_F(E) = (1 + e^{E/k_B T})^{-1}$ is the Fermi-Dirac distribution.
For $\delta \xi_\mathbf{k} = 0$, these reduce to
\begin{subequations}
\begin{gather}
    \langle 
    d^\dagger_{-\mathbf{k}} 
    d_{-\mathbf{k}}
    \rangle
    =
    \langle 
    \chi^\dagger_{\mathbf{k}, +} 
    \chi_{\mathbf{k}, +}
    \rangle
    =
    \frac{1}{2} \left( 1 - \frac{\bar{\xi}_\mathbf{k}}{E_\mathbf{k}}\right) + \frac{\bar{\xi}_\mathbf{k}}{E_\mathbf{k}} n_F(E_\mathbf{k}),
    \\
    \langle
    d_{-\mathbf{k}}
    \chi_{\mathbf{k}, +}
    \rangle
    =
    - \frac{\Delta^{(+)}(\mathbf{k})}{2E_\mathbf{k}}
    \tanh
    \left(
    \frac{E_\mathbf{k}}{2k_B T}
    \right).
\end{gather}
\end{subequations}
At zero temperature, the energy reduces to
\begin{align}
    E_\mathrm{GS}
    & \overset{T \rightarrow 0}{=}
    \sum_{\mathbf{k}}
    (\bar{\xi}_{\mathbf{k}} - E_\mathbf{k})
    -
    \frac{N |\Delta^{(+)}|^2}{V_0}
    +
    \frac{1}{2}
    \sum_{\mathbf{k}} \xi_{\mathbf{k}, -}^{(c)} 
\end{align}
The latter contribution $\frac{1}{2}
\sum_{\mathbf{k}} \xi_{\mathbf{k}, -}^{(c)}$
corresponds to the unpaired bands, which is identical to that of the normal state.

First, we differentiate with respect to $\Delta^{(+)}$ to find the energy minimum,
\begin{align}
    \left.
    \frac{\partial E_\mathrm{GS}}{\partial \Delta^{(+)*}}
    \right|_{T = 0}
    &= \nonumber
    -\sum_{\mathbf{k}}
    \frac{\partial E_\mathbf{k}}{\partial \Delta^{(+)*}}
    -
    \frac{N \Delta^{(+)}}{V_0}
    \\
    &=
    -\sum_{\mathbf{k}}
    \frac{\Delta^{(+)} |\phi(\mathbf{k})|^2}{2\sqrt{\bar{\xi}_\mathbf{k}^2 + |\Delta^{(+)} \phi(\mathbf{k})|^2 }}
    -
    \frac{N \Delta^{(+)}}{V_0}
\end{align}
Above, we have used the Fermi surface matching condition, $(\xi^{(c)}_{\mathbf{k}, +} - \xi_{-\mathbf{k}}^{(d)}) \approx 0$.
Any deviation would represent higher order corrections to the condensation energy.
As such, the minimum is satisfied under the following condition,
\begin{equation}
    1
    =
    - \frac{V_0}{N}
    \sum_{\mathbf{k}}
    \frac{|\phi(\mathbf{k})|^2}{2 E_\mathbf{k}},
\end{equation}
which is true for attractive interaction $V_0<0$.
Taking the second order derivative, we find
\begin{align}
    \left.
    \frac{\partial^2 E_\mathrm{GS}}{\partial \Delta^{(+)}\partial \Delta^{(+)*}}
    \right|_{T = 0}
    =
    \sum_\mathbf{k}
    \frac{|\phi(\mathbf{k})|^2}{2 E_{\mathbf{k}}}
    \left(
    \frac{|\Delta^{(+)}|^2 |\phi(\mathbf{k})|^2}{2 E_\mathbf{k}^2} - 1
    \right)
    -
    \frac{N}{V_0},
\end{align}
At the minimum, the second derivative is given by
\begin{align}
    \left.
    \frac{\partial^2 E_\mathrm{GS}}{\partial \Delta^{(+)}\partial \Delta^{(+)*}}
    \right|_{T = 0; \mathrm{min}}
    =
    \sum_\mathbf{k}
    \frac{|\Delta^{(+)}|^2 |\phi(\mathbf{k})|^4}{4 E_\mathbf{k}^{3}},
\end{align}
which
is positive definite.
As such, spinor pairing order between $c$ and $d$ fermions is energetically stable.

\section{Free energy analysis}

In this section, we comment on the stability of the pairing gap by an analysis of the free energy.
Consider the continuum model of a spinor superconductor described in Eqs.~(4) and (5) of the main text.
Without loss of generality, we examine the case when $\mathrm{FS}_d$ is matched with $\mathrm{FS}_{c, +}$ and $\boldsymbol{\Delta}(\mathbf{k}) = (\Delta_\uparrow(\mathbf{k}), 0)^\mathrm{T}$.
Suppose that $\Delta_\uparrow(\mathbf{k})$ may be decomposed into partial wave channels that all transform under rotation according to $m_j = 1/2$.
Regardless of the partial wave symmetry of $\Delta_\uparrow(\mathbf{k})$, the topology of the states at the Fermi surface dictates the pair monopole charge, $q_{pair}=-1/2$, which sets a lower bound to the partial wave channels of the pairing gap function.

\begin{figure}
    \centering
    \includegraphics[width=\linewidth]{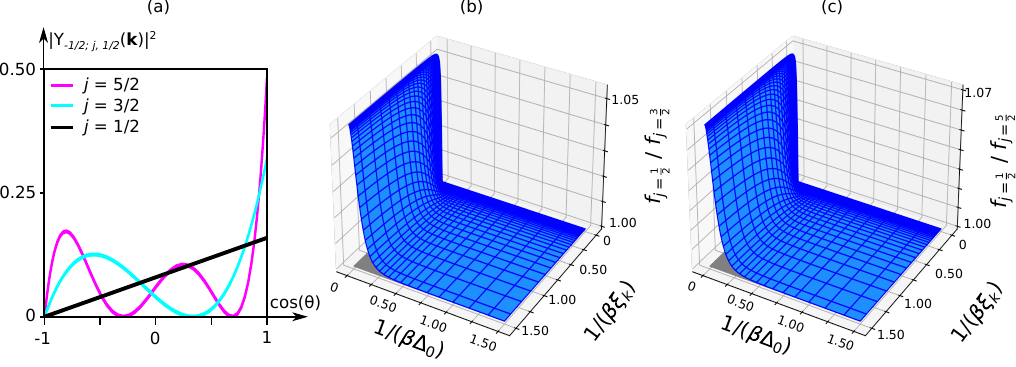}
    \caption{(a)~Angular distribution of the gap function 
    $|{Y}_{- \frac{1}{2}; j, \frac{1}{2}}(\hat{\mathbf{k}})|^2$ 
    corresponding to $q_{pair} = -1/2$ and $m_j = 1/2$.
    The magnitude of the gap function is shown for the $j = 1/2$, $3/2$, and $5/2$ partial wave channels in black, cyan, and magenta respectively.
    (b-c)~Ratio of free energy density $f_{j}(k)$ in Eq.~\eqref{free_energy_density} for (b)~pairing in the $j = 1/2$ and $j=3/2$ partial wave channels and (c)~pairing in the $j = 1/2$ and $j=5/2$ partial wave channels, assuming that the pairing channels have equal amplitude, $\Delta_0$.
    }
    \label{fig:pairing_gap_magnitude}
\end{figure}

After the mean-field decomposition and projection in Eq.~(6) of the main text, 
pairing wave amplitudes $\Delta_{j, m_j=\frac{1}{2}}$ can have nonvanishing amplitude.
As such, the pairing order is given as a superposition of monopole harmonics with the same conserved $m_j=1/2$ but in different partial wave channels, 
$\Delta^{(+)}(\mathbf{k}) = \sum_j \Delta_{j, \frac{1}{2}} Y_{-\frac{1}{2}; j, \frac{1}{2}}(\hat{\mathbf{k}})$ for
$j = n+1/2$ and $n = 0, 1,2,\cdots$.
In Fig.~\ref{fig:pairing_gap_magnitude}~(a), we show the angular distribution of the lowest three angular momentum pairing channels, $j=1/2$, $3/2$, and $5/2$, allowed by the symmetry of the system.

We now consider the stability of the different partial wave channels.
The free energy may be written as
\begin{math}
    F = (-1/\beta)\ln \mathcal{Z}
\end{math}
in which
\begin{math}
    \mathcal{Z} = {\sum_{n}}' 2\cos (\beta E_n)
\end{math}
is the partition function.
Here, $E_n$ are the BdG quasiparticle energies, $\beta = 1/ k_B T$, and the sum is taken over states with $E_n <0$.
Let us consider the case in which there is no spatial variation of the pairing order.
The free energy for the $j^\mathrm{th}$ partial wave channel is given by
\begin{equation}
    F_j = 
    \frac{V}{(2\pi)^3}
    \int \mathrm{d} k \int \mathrm{d} \phi \ k^2 f_j(k)
\end{equation}
in which
\begin{equation}
    f_j(k) = - \frac{1}{\beta}  \int \mathrm{d}(-\cos \theta)
    \ln 
    \left(
    2 \cosh\left( \frac{\beta}{2} \sqrt{|\Delta_{j, \frac{1}{2}} Y_{-\frac{1}{2}; j, \frac{1}{2}}|^2 + \epsilon_k^2}
    \right)
    \right)
    \label{free_energy_density}
\end{equation}
is the free-energy density for the partial wave channel $j$, with $\epsilon_k$ being the band dispersion.

In Fig.~\ref{fig:pairing_gap_magnitude}~(b), we show the ratio of the free energy densities for partial wave channels $j = 1/2$ and $3/2$, considering a general anisotropic pairing interaction so that each partial wave channel is equally weighted, $\Delta_{j=\frac{1}{2}, \frac{1}{2}} = \Delta_{j=\frac{3}{2}, \frac{1}{2}} = \Delta_0$.
The ratio $f_{\frac{1}{2}}/f_{\frac{3}{2}}$ is always greater than 1; hence, the pairing in the lower partial wave channel is favored energetically.
Similarly, if we consider the ratio $f_{\frac{1}{2}}/f_{\frac{5}{2}}$ for $\Delta_{j=\frac{1}{2}, \frac{1}{2}} = \Delta_{j=\frac{5}{2}, \frac{1}{2}} = \Delta_0$, the pairing in the $j=1/2$ channel is preferable to that in the $j=5/2$ channel for the same reason, as shown in Fig.~\ref{fig:pairing_gap_magnitude}~(c).
More generally, the lowest $j=1/2$ partial wave channel corresponds to the pairing channel with the least number of nodes.
Without the presence of additional symmetry-breaking terms, pairing in the $j = 1/2$ channel is the most energetically stable.

\section{Fractional Mermin-Ho relation}

In this appendix, we derive the fractional Mermin-Ho relation for the spinor superconducting system.
Consider a general spinor pairing order given by $\Delta(\mathbf{r}) = |\Delta(\mathbf{r})| e^{i \phi(\mathbf{r})} \eta(\mathbf{r})$, in which $\eta(\mathbf{r}) = (\eta_\uparrow(\mathbf{r}), \eta_\downarrow(\mathbf{r}))^\mathrm{T}$ is a normalized spinor transforming according to the fundamental spin-$1/2$ representation and satisfying $\eta^\dagger (\mathbf{r}) \eta (\mathbf{r}) = 1$.

The superfluid velocity is generally given by
\begin{equation}
    \mathbf{v}_{s} = -i \frac{\hbar}{m^*} \frac{1}{2} (\hat{\Delta}^{\dagger} \boldsymbol{\nabla} \hat{\Delta} - \mathrm{h.c.})
\end{equation}
in which $m^*$ is the mass of the Cooper pair.
The superfluid velocity of the spinor pairing order takes the form
\begin{equation}
    \mathbf{v}_{s} = 
    \frac{\hbar}{m^*}
    \Big(
    \boldsymbol{\nabla} \phi(\mathbf{r})
    -i
    \eta^*_\alpha \boldsymbol{\nabla} \eta_\alpha
    \Big),
\end{equation}
where summation is implied over repeated indices ($\alpha = \uparrow, \downarrow$).
As the U($1$) phase $\phi(\mathbf{r})$  varies smoothly over a simply connected region,
$\boldsymbol{\nabla} \times \boldsymbol{\nabla} \phi(\mathbf{r}) =0$.
However, the second term can contribute to the nonzero curl of the superfluid velocity,
\begin{equation}
    (\boldsymbol{\nabla} \times \mathbf{v}_s)_i
    =
    -i 
    \frac{\hbar}{m^*}
    (\boldsymbol{\nabla} \times \eta^*_\alpha \boldsymbol{\nabla} \eta_\alpha)_i
    =
    -i 
    \frac{\hbar}{m^*}
    \epsilon_{ijk} \partial_j \eta^*_\alpha \partial_k \eta_\alpha.
\end{equation}

Because $\eta(\mathbf{r})$ is normalized, it corresponds to a point on the three-sphere $S^3$.
We now use the Hopf map, $S^3 \rightarrow S^2$, to map the spinor $\eta = (\eta_\uparrow, \eta_\downarrow)^\mathrm{T}$ to a point $\hat{\mathbf{n}}$ on the two-sphere $S^2$ via~\cite{Stone2009SM}
\begin{equation}
    \hat{n}^i = \eta^\dagger \sigma^i \eta,
\end{equation}
where $\sigma^i$ is the Pauli matrix ($i = 1,2,3$).
Here, the components of the unit vector $\hat{\mathbf{n}}$ are given by
\begin{equation}
\begin{aligned}
    \hat{n}^1 = \eta^*_\uparrow \eta_\downarrow + \eta^*_\downarrow \eta_\uparrow;
    \hspace{2em}
    \hat{n}^2 = \frac{1}{i} (\eta^*_\uparrow \eta_\downarrow - \eta^*_\downarrow \eta_\uparrow);
    \hspace{2em}
    \hat{n}^3 = |\eta_\uparrow|^2 - |\eta_\downarrow|^2.
\end{aligned}
\label{n_components}
\end{equation}

Using the antisymmetry of $\epsilon_{abc}$ and the expression of $\hat{\mathbf{n}}$ in terms of the spinor components in Eq.~\eqref{n_components}, it can be shown that
\begin{align}
    \epsilon_{abc} \hat{n}^a \partial_j \hat{n}^b \partial_k \hat{n}^c
    = \nonumber
    \frac{2}{i}
    \bigg\{
    &
    \partial_k \left( | \eta_\uparrow|^2 - |\eta_\downarrow|^2 \right)
    \Big(
    - \left( \eta_\uparrow^* \eta_\downarrow \right) \partial_j \left( \eta_\downarrow^* \eta_\uparrow \right) 
    +
    \left(\eta_\downarrow^* \eta_\uparrow \right) \partial_j \left( \eta_\uparrow^* \eta_\downarrow \right)
    \Big)
    \\ \nonumber
    &-
    \partial_j \left( | \eta_\uparrow|^2 - |\eta_\downarrow|^2 \right )
    \Big(
    -\left( \eta_\uparrow^* \eta_\downarrow \right) \partial_k \left( \eta_\downarrow^* \eta_\uparrow \right)
    +
    \left( \eta_\downarrow^* \eta_\uparrow \right) \partial_k \left( \eta_\uparrow^* \eta_\downarrow \right)
    \Big)
    \\
    & 
    + \left( | \eta_\uparrow|^2 - |\eta_\downarrow|^2 \right) 
    \Big( - \partial_j \left( \eta_\uparrow^* \eta_\downarrow \right) \partial_k \left( \eta_\downarrow^* \eta_\uparrow \right)
    +
    \partial_j \left( \eta_\downarrow^* \eta_\uparrow \right) \partial_k \left( \eta_\uparrow^* \eta_\downarrow \right)
    \Big)
    \bigg\}
    \nonumber
    \\
    = \nonumber
    \frac{2}{i}
    \Bigg\{
    &
    \eta_\uparrow^* \eta_\downarrow \Big[
    (\partial_j \eta_\uparrow)(\partial_k \eta_\downarrow^*)
    -
    (\partial_k \eta_\uparrow)(\partial_j \eta_\downarrow^*)
    \Big]
    +
    \eta_\uparrow \eta_\downarrow^*
    \Big[
    (\partial_j \eta_\downarrow)(\partial_k \eta_\uparrow^*)
    -
    (\partial_k \eta_\downarrow)(\partial_j \eta_\uparrow^*)
    \Big]
    \\
    & + |\eta_\uparrow|^2
    \Big[
    (\partial_j \eta_\downarrow^*) (\partial_k \eta_\downarrow)
    -
    (\partial_k \eta_\downarrow^*) (\partial_j \eta_\downarrow)
    \Big]
    +
    |\eta_\downarrow|^2
    \Big[
    (\partial_j \eta_\uparrow^*) (\partial_k \eta_\uparrow)
    -
    (\partial_k \eta_\uparrow^*) (\partial_j \eta_\uparrow)
    \Big]
    \Bigg\}.
    \label{intermediate_stepA}
\end{align}
Above, $a,b,c$ and $j, k$ take values $1,2,3$, with summation still implied over repeated indices.
Additionally, as the spinor is normalized, it follows that
\begin{align}
    0 &= \partial_j (|\eta_\uparrow|^2 + |\eta_\downarrow|^2) = \partial_k (|\eta_\uparrow|^2 + |\eta_\downarrow|^2).
\end{align}
This may be rewritten as
\begin{subequations}
\begin{align}
    (\partial_j \eta_\uparrow^*) \eta_\uparrow
    +
    (\partial_j \eta_\downarrow^*) \eta_\downarrow
    =
    -
    \Big[
    (\partial_j \eta_\uparrow) \eta_\uparrow^*
    +
    (\partial_j \eta_\downarrow) \eta_\downarrow^*
    \Big],
    \\
    (\partial_k \eta_\uparrow) \eta_\uparrow^*
    +
    (\partial_k \eta_\downarrow) \eta_\downarrow^*
    =
    -
    \Big[
    (\partial_k \eta_\uparrow^*) \eta_\uparrow
    +
    (\partial_k \eta_\downarrow^*) \eta_\downarrow
    \Big].
\end{align}
\end{subequations}
Multiplying the two equations and rearranging terms, we arrive to the identity
\begin{equation}
\begin{aligned}
    |\eta_\uparrow|^2
    \Big[
    (\partial_j \eta^*_\uparrow)
    (\partial_k \eta_\uparrow)
    -
    (\partial_k \eta^*_\uparrow)
    (\partial_j \eta_\uparrow)
    \Big]
    +
    |\eta_\downarrow|^2
    \Big[
    (\partial_j \eta^*_\downarrow)
    (\partial_k \eta_\downarrow)
    -
    (\partial_k \eta^*_\downarrow)
    (\partial_j \eta_\downarrow)
    \Big]
    \\
    =
    \eta_\uparrow \eta^*_\downarrow
    \Big[
    (\partial_k \eta^*_\uparrow)
    (\partial_j \eta_\downarrow)
    -
    (\partial_j \eta^*_\uparrow)
    (\partial_k \eta_\downarrow)
    \Big]
    +
    \eta^*_\uparrow \eta_\downarrow
    \Big[
    (\partial_k \eta^*_\downarrow)
    (\partial_j \eta_\uparrow)
    -
    (\partial_j \eta^*_\downarrow)
    (\partial_k \eta_\uparrow)
    \Big].
    \label{normalization_identity}
\end{aligned}
\end{equation}

With Eq.~\eqref{normalization_identity} and the expression in Eq.~\eqref{intermediate_stepA},
it follows that~\cite{Wilczek1983SM, Stone2009SM}
\begin{equation}
    \epsilon_{abc} \hat{n}^a \partial_j \hat{n}^b \partial_k \hat{n}^c
    =
    \frac{2}{i}
    \Big[
    (\partial_j \eta^*_\alpha)(\partial_k \eta_\alpha)
    -
    (\partial_k \eta^*_\alpha)(\partial_j \eta_\alpha)
    \Big].
\end{equation}
The left-hand side is related to the topological current density in the $\mathrm{O}(3)$ nonlinear $\sigma$ model, and the right-hand side, in analogy to Ref.~\citenum{Wilczek1983SM}, corresponds to the curl of a gauge potential, $A_j = i \eta^\dagger \partial_j \eta$.
Here, in contrast to Ref.~\citenum{Wilczek1983SM} in which the gauge potential is manufactured, this term arises naturally as the superfluid velocity.
Considering the antisymmetry of $\epsilon_{ijk}$ and summing over $j$ and $k$, the curl of the superfluid velocity can be rewritten as
\begin{align}
    (\boldsymbol{\nabla} \times \mathbf{v}_s)_i
    &= 
    -i 
    \frac{\hbar}{m^*}
    \epsilon_{ijk} \partial_j \eta^*_\alpha \partial_k \eta_\alpha
    =
    \frac{1}{4}
    \frac{\hbar}{m^*}
    \epsilon_{abc} \epsilon_{ijk} \hat{n}^a \partial_j \hat{n}^b \partial_k \hat{n}^c.
\end{align}
Comparing to the celebrated Mermin-Ho relation~\cite{Mermin1976SM, Kamien2002SM},
$(\boldsymbol{\nabla} \times \mathbf{v}_s)_i = \frac{1}{2} \frac{\hbar}{m^*}
\epsilon_{abc} \epsilon_{ijk} \hat{l}^a \partial_j \hat{l}^b \partial_k \hat{l}^c$, the $\hat{n}$-vector here now plays the role of the $\hat{l}$-vector.
The additional overall factor of $1/2$ indicates that the spinor superconducting system obeys a fractional Mermin-Ho relation.



\begin{thebibliography}{59}%
\makeatletter
\providecommand \@ifxundefined [1]{%
 \@ifx{#1\undefined}
}%
\providecommand \@ifnum [1]{%
 \ifnum #1\expandafter \@firstoftwo
 \else \expandafter \@secondoftwo
 \fi
}%
\providecommand \@ifx [1]{%
 \ifx #1\expandafter \@firstoftwo
 \else \expandafter \@secondoftwo
 \fi
}%
\providecommand \natexlab [1]{#1}%
\providecommand \enquote  [1]{``#1''}%
\providecommand \bibnamefont  [1]{#1}%
\providecommand \bibfnamefont [1]{#1}%
\providecommand \citenamefont [1]{#1}%
\providecommand \href@noop [0]{\@secondoftwo}%
\providecommand \href [0]{\begingroup \@sanitize@url \@href}%
\providecommand \@href[1]{\@@startlink{#1}\@@href}%
\providecommand \@@href[1]{\endgroup#1\@@endlink}%
\providecommand \@sanitize@url [0]{\catcode `\\12\catcode `\$12\catcode
  `\&12\catcode `\#12\catcode `\^12\catcode `\_12\catcode `\%12\relax}%
\providecommand \@@startlink[1]{}%
\providecommand \@@endlink[0]{}%
\providecommand \url  [0]{\begingroup\@sanitize@url \@url }%
\providecommand \@url [1]{\endgroup\@href {#1}{\urlprefix }}%
\providecommand \urlprefix  [0]{URL }%
\providecommand \Eprint [0]{\href }%
\providecommand \doibase [0]{http://dx.doi.org/}%
\providecommand \selectlanguage [0]{\@gobble}%
\providecommand \bibinfo  [0]{\@secondoftwo}%
\providecommand \bibfield  [0]{\@secondoftwo}%
\providecommand \translation [1]{[#1]}%
\providecommand \BibitemOpen [0]{}%
\providecommand \bibitemStop [0]{}%
\providecommand \bibitemNoStop [0]{.\EOS\space}%
\providecommand \EOS [0]{\spacefactor3000\relax}%
\providecommand \BibitemShut  [1]{\csname bibitem#1\endcsname}%
\let\auto@bib@innerbib\@empty
\bibitem [{\citenamefont {Anderson}\ and\ \citenamefont
  {Morel}(1961)}]{Anderson1961}%
  \BibitemOpen
  \bibfield  {author} {\bibinfo {author} {\bibfnamefont {PW}~\bibnamefont
  {Anderson}}\ and\ \bibinfo {author} {\bibfnamefont {P.}~\bibnamefont
  {Morel}},\ }\bibfield  {title} {\enquote {\bibinfo {title} {Generalized
  bardeen-cooper-schrieffer states and the proposed low-temperature phase of
  liquid {He}$^3$},}\ }\href@noop {} {\bibfield  {journal} {\bibinfo  {journal}
  {Phys. Rev.}\ }\textbf {\bibinfo {volume} {123}},\ \bibinfo {pages} {1911}
  (\bibinfo {year} {1961})}\BibitemShut {NoStop}%
\bibitem [{\citenamefont {Balian}\ and\ \citenamefont
  {Werthamer}(1963)}]{Balian1963}%
  \BibitemOpen
  \bibfield  {author} {\bibinfo {author} {\bibfnamefont {R.}~\bibnamefont
  {Balian}}\ and\ \bibinfo {author} {\bibfnamefont {NR}~\bibnamefont
  {Werthamer}},\ }\bibfield  {title} {\enquote {\bibinfo {title}
  {Superconductivity with pairs in a relative p wave},}\ }\href@noop {}
  {\bibfield  {journal} {\bibinfo  {journal} {Phys. Rev.}\ }\textbf {\bibinfo
  {volume} {131}},\ \bibinfo {pages} {1553} (\bibinfo {year}
  {1963})}\BibitemShut {NoStop}%
\bibitem [{\citenamefont {Leggett}(1975)}]{Leggett1975}%
  \BibitemOpen
  \bibfield  {author} {\bibinfo {author} {\bibfnamefont {Anthony~J.}\
  \bibnamefont {Leggett}},\ }\bibfield  {title} {\enquote {\bibinfo {title} {A
  theoretical description of the new phases of liquid $^{3}\mathrm{He}$},}\
  }\href {\doibase 10.1103/RevModPhys.47.331} {\bibfield  {journal} {\bibinfo
  {journal} {Rev. Mod. Phys.}\ }\textbf {\bibinfo {volume} {47}},\ \bibinfo
  {pages} {331--414} (\bibinfo {year} {1975})}\BibitemShut {NoStop}%
\bibitem [{\citenamefont {Volovik}(2003)}]{Volovik2003}%
  \BibitemOpen
  \bibfield  {author} {\bibinfo {author} {\bibfnamefont {Grigory~E}\
  \bibnamefont {Volovik}},\ }\href@noop {} {\emph {\bibinfo {title} {The
  universe in a helium droplet}}},\ Vol.\ \bibinfo {volume} {117}\ (\bibinfo
  {publisher} {Oxford University Press},\ \bibinfo {year} {2003})\BibitemShut
  {NoStop}%
\bibitem [{\citenamefont {Tsuei}\ \emph {et~al.}(1994)\citenamefont {Tsuei},
  \citenamefont {Kirtley}, \citenamefont {Chi}, \citenamefont {Yu-Jahnes},
  \citenamefont {Gupta}, \citenamefont {Shaw}, \citenamefont {Sun},\ and\
  \citenamefont {Ketchen}}]{Tsuei1994}%
  \BibitemOpen
  \bibfield  {author} {\bibinfo {author} {\bibfnamefont {C.~C.}\ \bibnamefont
  {Tsuei}}, \bibinfo {author} {\bibfnamefont {J.~R.}\ \bibnamefont {Kirtley}},
  \bibinfo {author} {\bibfnamefont {C.~C.}\ \bibnamefont {Chi}}, \bibinfo
  {author} {\bibfnamefont {Lock~See}\ \bibnamefont {Yu-Jahnes}}, \bibinfo
  {author} {\bibfnamefont {A.}~\bibnamefont {Gupta}}, \bibinfo {author}
  {\bibfnamefont {T.}~\bibnamefont {Shaw}}, \bibinfo {author} {\bibfnamefont
  {J.~Z.}\ \bibnamefont {Sun}}, \ and\ \bibinfo {author} {\bibfnamefont
  {M.~B.}\ \bibnamefont {Ketchen}},\ }\bibfield  {title} {\enquote {\bibinfo
  {title} {Pairing symmetry and flux quantization in a tricrystal
  superconducting ring of {YBa$_2$Cu$_3$O$_7$}},}\ }\href {\doibase
  10.1103/PhysRevLett.73.593} {\bibfield  {journal} {\bibinfo  {journal} {Phys.
  Rev. Lett.}\ }\textbf {\bibinfo {volume} {73}},\ \bibinfo {pages} {593--596}
  (\bibinfo {year} {1994})}\BibitemShut {NoStop}%
\bibitem [{\citenamefont {Van~Harlingen}(1995)}]{VanHarlingen1995}%
  \BibitemOpen
  \bibfield  {author} {\bibinfo {author} {\bibfnamefont {D.~J.}\ \bibnamefont
  {Van~Harlingen}},\ }\bibfield  {title} {\enquote {\bibinfo {title}
  {Phase-sensitive tests of the symmetry of the pairing state in the
  high-temperature superconductors\char22{}evidence for ${d}_{{x}^{2}-{y}^{2}}$
  symmetry},}\ }\href {\doibase 10.1103/RevModPhys.67.515} {\bibfield
  {journal} {\bibinfo  {journal} {Rev. Mod. Phys.}\ }\textbf {\bibinfo {volume}
  {67}},\ \bibinfo {pages} {515--535} (\bibinfo {year} {1995})}\BibitemShut
  {NoStop}%
\bibitem [{\citenamefont {Tsuei}\ and\ \citenamefont
  {Kirtley}(2000)}]{Tsuei2000}%
  \BibitemOpen
  \bibfield  {author} {\bibinfo {author} {\bibfnamefont {C.~C.}\ \bibnamefont
  {Tsuei}}\ and\ \bibinfo {author} {\bibfnamefont {J.~R.}\ \bibnamefont
  {Kirtley}},\ }\bibfield  {title} {\enquote {\bibinfo {title} {Pairing
  symmetry in cuprate superconductors},}\ }\href {\doibase
  10.1103/RevModPhys.72.969} {\bibfield  {journal} {\bibinfo  {journal} {Rev.
  Mod. Phys.}\ }\textbf {\bibinfo {volume} {72}},\ \bibinfo {pages} {969--1016}
  (\bibinfo {year} {2000})}\BibitemShut {NoStop}%
\bibitem [{\citenamefont {Stewart}(2011)}]{Stewart2011}%
  \BibitemOpen
  \bibfield  {author} {\bibinfo {author} {\bibfnamefont {G.~R.}\ \bibnamefont
  {Stewart}},\ }\bibfield  {title} {\enquote {\bibinfo {title}
  {Superconductivity in iron compounds},}\ }\href {\doibase
  10.1103/RevModPhys.83.1589} {\bibfield  {journal} {\bibinfo  {journal} {Rev.
  Mod. Phys.}\ }\textbf {\bibinfo {volume} {83}},\ \bibinfo {pages}
  {1589--1652} (\bibinfo {year} {2011})}\BibitemShut {NoStop}%
\bibitem [{\citenamefont {Dai}(2015)}]{Dai2015}%
  \BibitemOpen
  \bibfield  {author} {\bibinfo {author} {\bibfnamefont {Pengcheng}\
  \bibnamefont {Dai}},\ }\bibfield  {title} {\enquote {\bibinfo {title}
  {Antiferromagnetic order and spin dynamics in iron-based superconductors},}\
  }\href {\doibase 10.1103/RevModPhys.87.855} {\bibfield  {journal} {\bibinfo
  {journal} {Rev. Mod. Phys.}\ }\textbf {\bibinfo {volume} {87}},\ \bibinfo
  {pages} {855--896} (\bibinfo {year} {2015})}\BibitemShut {NoStop}%
\bibitem [{\citenamefont {Haldane}(1988)}]{Haldane1988}%
  \BibitemOpen
  \bibfield  {author} {\bibinfo {author} {\bibfnamefont {F.~D.~M.}\
  \bibnamefont {Haldane}},\ }\bibfield  {title} {\enquote {\bibinfo {title}
  {Model for a quantum hall effect without landau levels: Condensed-matter
  realization of the "parity anomaly"},}\ }\href {\doibase
  10.1s103/PhysRevLett.61.2015} {\bibfield  {journal} {\bibinfo  {journal}
  {Phys. Rev. Lett.}\ }\textbf {\bibinfo {volume} {61}},\ \bibinfo {pages}
  {2015--2018} (\bibinfo {year} {1988})}\BibitemShut {NoStop}%
\bibitem [{\citenamefont {King-Smith}\ and\ \citenamefont
  {Vanderbilt}(1993)}]{King-Smith1993}%
  \BibitemOpen
  \bibfield  {author} {\bibinfo {author} {\bibfnamefont {RD}~\bibnamefont
  {King-Smith}}\ and\ \bibinfo {author} {\bibfnamefont {D.}~\bibnamefont
  {Vanderbilt}},\ }\bibfield  {title} {\enquote {\bibinfo {title} {{Theory of
  polarization of crystalline solids}},}\ }\href@noop {} {\bibfield  {journal}
  {\bibinfo  {journal} {Phys. Rev. B}\ }\textbf {\bibinfo {volume} {47}},\
  \bibinfo {pages} {1651--1654} (\bibinfo {year} {1993})}\BibitemShut {NoStop}%
\bibitem [{\citenamefont {Kane}\ and\ \citenamefont {Mele}(2005)}]{Kane2005a}%
  \BibitemOpen
  \bibfield  {author} {\bibinfo {author} {\bibfnamefont {C.~L.}\ \bibnamefont
  {Kane}}\ and\ \bibinfo {author} {\bibfnamefont {E.~J.}\ \bibnamefont
  {Mele}},\ }\bibfield  {title} {\enquote {\bibinfo {title} {${Z}_{2}$
  topological order and the quantum spin hall effect},}\ }\href {\doibase
  10.1103/PhysRevLett.95.146802} {\bibfield  {journal} {\bibinfo  {journal}
  {Phys. Rev. Lett.}\ }\textbf {\bibinfo {volume} {95}},\ \bibinfo {pages}
  {146802} (\bibinfo {year} {2005})}\BibitemShut {NoStop}%
\bibitem [{\citenamefont {Fu}\ \emph {et~al.}(2007)\citenamefont {Fu},
  \citenamefont {Kane},\ and\ \citenamefont {Mele}}]{Fu2007a}%
  \BibitemOpen
  \bibfield  {author} {\bibinfo {author} {\bibfnamefont {Liang}\ \bibnamefont
  {Fu}}, \bibinfo {author} {\bibfnamefont {C.~L.}\ \bibnamefont {Kane}}, \ and\
  \bibinfo {author} {\bibfnamefont {E.~J.}\ \bibnamefont {Mele}},\ }\bibfield
  {title} {\enquote {\bibinfo {title} {Topological insulators in three
  dimensions},}\ }\href {\doibase 10.1103/PhysRevLett.98.106803} {\bibfield
  {journal} {\bibinfo  {journal} {Phys. Rev. Lett.}\ }\textbf {\bibinfo
  {volume} {98}},\ \bibinfo {pages} {106803} (\bibinfo {year}
  {2007})}\BibitemShut {NoStop}%
\bibitem [{\citenamefont {Xiao}\ \emph {et~al.}(2010)\citenamefont {Xiao},
  \citenamefont {Chang},\ and\ \citenamefont {Niu}}]{Xiao2010}%
  \BibitemOpen
  \bibfield  {author} {\bibinfo {author} {\bibfnamefont {D.}~\bibnamefont
  {Xiao}}, \bibinfo {author} {\bibfnamefont {M.C.}\ \bibnamefont {Chang}}, \
  and\ \bibinfo {author} {\bibfnamefont {Q.}~\bibnamefont {Niu}},\ }\bibfield
  {title} {\enquote {\bibinfo {title} {{Berry phase effects on electronic
  properties}},}\ }\href@noop {} {\bibfield  {journal} {\bibinfo  {journal}
  {Rev. Mod. Phys.}\ }\textbf {\bibinfo {volume} {82}},\ \bibinfo {pages}
  {1959--2007} (\bibinfo {year} {2010})}\BibitemShut {NoStop}%
\bibitem [{\citenamefont {Yu}\ \emph {et~al.}(2010)\citenamefont {Yu},
  \citenamefont {Zhang}, \citenamefont {Zhang}, \citenamefont {Zhang},
  \citenamefont {Dai},\ and\ \citenamefont {Fang}}]{Yu2010}%
  \BibitemOpen
  \bibfield  {author} {\bibinfo {author} {\bibfnamefont {Rui}\ \bibnamefont
  {Yu}}, \bibinfo {author} {\bibfnamefont {Wei}\ \bibnamefont {Zhang}},
  \bibinfo {author} {\bibfnamefont {Hai-Jun}\ \bibnamefont {Zhang}}, \bibinfo
  {author} {\bibfnamefont {Shou-Cheng}\ \bibnamefont {Zhang}}, \bibinfo
  {author} {\bibfnamefont {Xi}~\bibnamefont {Dai}}, \ and\ \bibinfo {author}
  {\bibfnamefont {Zhong}\ \bibnamefont {Fang}},\ }\bibfield  {title} {\enquote
  {\bibinfo {title} {Quantized anomalous hall effect in magnetic topological
  insulators},}\ }\href {\doibase 10.1126/science.1187485} {\bibfield
  {journal} {\bibinfo  {journal} {Science}\ }\textbf {\bibinfo {volume}
  {329}},\ \bibinfo {pages} {61--64} (\bibinfo {year} {2010})}\BibitemShut
  {NoStop}%
\bibitem [{\citenamefont {Chang}\ \emph {et~al.}(2013)\citenamefont {Chang},
  \citenamefont {Zhang}, \citenamefont {Feng}, \citenamefont {Shen},
  \citenamefont {Zhang}, \citenamefont {Guo}, \citenamefont {Li}, \citenamefont
  {Ou}, \citenamefont {Wei}, \citenamefont {Wang}, \citenamefont {Ji},
  \citenamefont {Feng}, \citenamefont {Ji}, \citenamefont {Chen}, \citenamefont
  {Jia}, \citenamefont {Dai}, \citenamefont {Fang}, \citenamefont {Zhang},
  \citenamefont {He}, \citenamefont {Wang}, \citenamefont {Lu}, \citenamefont
  {Ma},\ and\ \citenamefont {Xue}}]{Chang2013}%
  \BibitemOpen
  \bibfield  {author} {\bibinfo {author} {\bibfnamefont {Cui-Zu}\ \bibnamefont
  {Chang}}, \bibinfo {author} {\bibfnamefont {Jinsong}\ \bibnamefont {Zhang}},
  \bibinfo {author} {\bibfnamefont {Xiao}\ \bibnamefont {Feng}}, \bibinfo
  {author} {\bibfnamefont {Jie}\ \bibnamefont {Shen}}, \bibinfo {author}
  {\bibfnamefont {Zuocheng}\ \bibnamefont {Zhang}}, \bibinfo {author}
  {\bibfnamefont {Minghua}\ \bibnamefont {Guo}}, \bibinfo {author}
  {\bibfnamefont {Kang}\ \bibnamefont {Li}}, \bibinfo {author} {\bibfnamefont
  {Yunbo}\ \bibnamefont {Ou}}, \bibinfo {author} {\bibfnamefont {Pang}\
  \bibnamefont {Wei}}, \bibinfo {author} {\bibfnamefont {Li-Li}\ \bibnamefont
  {Wang}}, \bibinfo {author} {\bibfnamefont {Zhong-Qing}\ \bibnamefont {Ji}},
  \bibinfo {author} {\bibfnamefont {Yang}\ \bibnamefont {Feng}}, \bibinfo
  {author} {\bibfnamefont {Shuaihua}\ \bibnamefont {Ji}}, \bibinfo {author}
  {\bibfnamefont {Xi}~\bibnamefont {Chen}}, \bibinfo {author} {\bibfnamefont
  {Jinfeng}\ \bibnamefont {Jia}}, \bibinfo {author} {\bibfnamefont
  {Xi}~\bibnamefont {Dai}}, \bibinfo {author} {\bibfnamefont {Zhong}\
  \bibnamefont {Fang}}, \bibinfo {author} {\bibfnamefont {Shou-Cheng}\
  \bibnamefont {Zhang}}, \bibinfo {author} {\bibfnamefont {Ke}~\bibnamefont
  {He}}, \bibinfo {author} {\bibfnamefont {Yayu}\ \bibnamefont {Wang}},
  \bibinfo {author} {\bibfnamefont {Li}~\bibnamefont {Lu}}, \bibinfo {author}
  {\bibfnamefont {Xu-Cun}\ \bibnamefont {Ma}}, \ and\ \bibinfo {author}
  {\bibfnamefont {Qi-Kun}\ \bibnamefont {Xue}},\ }\bibfield  {title} {\enquote
  {\bibinfo {title} {Experimental observation of the quantum anomalous hall
  effect in a magnetic topological insulator},}\ }\href {\doibase
  10.1126/science.1234414} {\bibfield  {journal} {\bibinfo  {journal}
  {Science}\ }\textbf {\bibinfo {volume} {340}},\ \bibinfo {pages} {167--170}
  (\bibinfo {year} {2013})}\BibitemShut {NoStop}%
\bibitem [{\citenamefont {{Haldane}}(2014)}]{Haldane2014}%
  \BibitemOpen
  \bibfield  {author} {\bibinfo {author} {\bibfnamefont {F.~D.~M.}\
  \bibnamefont {{Haldane}}},\ }\bibfield  {title} {\enquote {\bibinfo {title}
  {Attachment of surface ''fermi arcs'' to the bulk fermi surface:
  ''fermi-level plumbing'' in topological metals},}\ }\href
  {https://arxiv.org/abs/1401.0529} {\bibfield  {journal} {\bibinfo  {journal}
  {arXiv:1401.0529}\ } (\bibinfo {year} {2014})}\BibitemShut {NoStop}%
\bibitem [{\citenamefont {Liu}\ \emph {et~al.}(2016)\citenamefont {Liu},
  \citenamefont {Yang}, \citenamefont {Wu}, \citenamefont {Shekhar},
  \citenamefont {Jiang}, \citenamefont {Yang}, \citenamefont {Zhang},
  \citenamefont {Mo}, \citenamefont {Hussain}, \citenamefont {Yan},
  \citenamefont {Felser},\ and\ \citenamefont {Chen}}]{Liu2016}%
  \BibitemOpen
  \bibfield  {author} {\bibinfo {author} {\bibfnamefont {ZK}~\bibnamefont
  {Liu}}, \bibinfo {author} {\bibfnamefont {LX}~\bibnamefont {Yang}}, \bibinfo
  {author} {\bibfnamefont {SC}~\bibnamefont {Wu}}, \bibinfo {author}
  {\bibfnamefont {C}~\bibnamefont {Shekhar}}, \bibinfo {author} {\bibfnamefont
  {J}~\bibnamefont {Jiang}}, \bibinfo {author} {\bibfnamefont {HF}~\bibnamefont
  {Yang}}, \bibinfo {author} {\bibfnamefont {Y}~\bibnamefont {Zhang}}, \bibinfo
  {author} {\bibfnamefont {SK}~\bibnamefont {Mo}}, \bibinfo {author}
  {\bibfnamefont {Z}~\bibnamefont {Hussain}}, \bibinfo {author} {\bibfnamefont
  {B}~\bibnamefont {Yan}}, \bibinfo {author} {\bibfnamefont {C}~\bibnamefont
  {Felser}}, \ and\ \bibinfo {author} {\bibfnamefont {YL}~\bibnamefont
  {Chen}},\ }\bibfield  {title} {\enquote {\bibinfo {title} {Observation of
  unusual topological surface states in half-heusler compounds {LnPtBi}
  ({Ln=Lu,Y})},}\ }\href {\doibase 10.1038/ncomms12924} {\bibfield  {journal}
  {\bibinfo  {journal} {Nat. Commun.}\ }\textbf {\bibinfo {volume} {7}},\
  \bibinfo {pages} {12924} (\bibinfo {year} {2016})}\BibitemShut {NoStop}%
\bibitem [{\citenamefont {Murakami}(2007)}]{Murakami2007}%
  \BibitemOpen
  \bibfield  {author} {\bibinfo {author} {\bibfnamefont {Shuichi}\ \bibnamefont
  {Murakami}},\ }\bibfield  {title} {\enquote {\bibinfo {title} {Phase
  transition between the quantum spin hall and insulator phases in 3d:
  emergence of a topological gapless phase},}\ }\href
  {http://stacks.iop.org/1367-2630/9/i=9/a=356} {\bibfield  {journal} {\bibinfo
   {journal} {New J. Phys.}\ }\textbf {\bibinfo {volume} {9}},\ \bibinfo
  {pages} {356} (\bibinfo {year} {2007})}\BibitemShut {NoStop}%
\bibitem [{\citenamefont {Wan}\ \emph {et~al.}(2011)\citenamefont {Wan},
  \citenamefont {Turner}, \citenamefont {Vishwanath},\ and\ \citenamefont
  {Savrasov}}]{Wan2011}%
  \BibitemOpen
  \bibfield  {author} {\bibinfo {author} {\bibfnamefont {Xiangang}\
  \bibnamefont {Wan}}, \bibinfo {author} {\bibfnamefont {Ari~M.}\ \bibnamefont
  {Turner}}, \bibinfo {author} {\bibfnamefont {Ashvin}\ \bibnamefont
  {Vishwanath}}, \ and\ \bibinfo {author} {\bibfnamefont {Sergey~Y.}\
  \bibnamefont {Savrasov}},\ }\bibfield  {title} {\enquote {\bibinfo {title}
  {Topological semimetal and fermi-arc surface states in the electronic
  structure of pyrochlore iridates},}\ }\href {\doibase
  10.1103/PhysRevB.83.205101} {\bibfield  {journal} {\bibinfo  {journal} {Phys.
  Rev. B}\ }\textbf {\bibinfo {volume} {83}},\ \bibinfo {pages} {205101}
  (\bibinfo {year} {2011})}\BibitemShut {NoStop}%
\bibitem [{\citenamefont {Xu}\ \emph {et~al.}(2011)\citenamefont {Xu},
  \citenamefont {Weng}, \citenamefont {Wang}, \citenamefont {Dai},\ and\
  \citenamefont {Fang}}]{Xu2011}%
  \BibitemOpen
  \bibfield  {author} {\bibinfo {author} {\bibfnamefont {Gang}\ \bibnamefont
  {Xu}}, \bibinfo {author} {\bibfnamefont {Hongming}\ \bibnamefont {Weng}},
  \bibinfo {author} {\bibfnamefont {Zhijun}\ \bibnamefont {Wang}}, \bibinfo
  {author} {\bibfnamefont {Xi}~\bibnamefont {Dai}}, \ and\ \bibinfo {author}
  {\bibfnamefont {Zhong}\ \bibnamefont {Fang}},\ }\bibfield  {title} {\enquote
  {\bibinfo {title} {Chern semimetal and the quantized anomalous hall effect in
  {HgCr}$_{2}${Se}$_{4}$},}\ }\href {\doibase 10.1103/PhysRevLett.107.186806}
  {\bibfield  {journal} {\bibinfo  {journal} {Phys. Rev. Lett.}\ }\textbf
  {\bibinfo {volume} {107}},\ \bibinfo {pages} {186806} (\bibinfo {year}
  {2011})}\BibitemShut {NoStop}%
\bibitem [{\citenamefont {Yang}\ \emph {et~al.}(2011)\citenamefont {Yang},
  \citenamefont {Lu},\ and\ \citenamefont {Ran}}]{Yang2011}%
  \BibitemOpen
  \bibfield  {author} {\bibinfo {author} {\bibfnamefont {Kai-Yu}\ \bibnamefont
  {Yang}}, \bibinfo {author} {\bibfnamefont {Yuan-Ming}\ \bibnamefont {Lu}}, \
  and\ \bibinfo {author} {\bibfnamefont {Ying}\ \bibnamefont {Ran}},\
  }\bibfield  {title} {\enquote {\bibinfo {title} {Quantum hall effects in a
  weyl semimetal: Possible application in pyrochlore iridates},}\ }\href
  {\doibase 10.1103/PhysRevB.84.075129} {\bibfield  {journal} {\bibinfo
  {journal} {Phys. Rev. B}\ }\textbf {\bibinfo {volume} {84}},\ \bibinfo
  {pages} {075129} (\bibinfo {year} {2011})}\BibitemShut {NoStop}%
\bibitem [{\citenamefont {Burkov}\ and\ \citenamefont
  {Balents}(2011)}]{Burkov2011}%
  \BibitemOpen
  \bibfield  {author} {\bibinfo {author} {\bibfnamefont {A.~A.}\ \bibnamefont
  {Burkov}}\ and\ \bibinfo {author} {\bibfnamefont {Leon}\ \bibnamefont
  {Balents}},\ }\bibfield  {title} {\enquote {\bibinfo {title} {Weyl semimetal
  in a topological insulator multilayer},}\ }\href {\doibase
  10.1103/PhysRevLett.107.127205} {\bibfield  {journal} {\bibinfo  {journal}
  {Phys. Rev. Lett.}\ }\textbf {\bibinfo {volume} {107}},\ \bibinfo {pages}
  {127205} (\bibinfo {year} {2011})}\BibitemShut {NoStop}%
\bibitem [{\citenamefont {Witczak-Krempa}\ and\ \citenamefont
  {Kim}(2012)}]{Witczak-Krempa2012}%
  \BibitemOpen
  \bibfield  {author} {\bibinfo {author} {\bibfnamefont {William}\ \bibnamefont
  {Witczak-Krempa}}\ and\ \bibinfo {author} {\bibfnamefont {Yong~Baek}\
  \bibnamefont {Kim}},\ }\bibfield  {title} {\enquote {\bibinfo {title}
  {Topological and magnetic phases of interacting electrons in the pyrochlore
  iridates},}\ }\href {\doibase 10.1103/PhysRevB.85.045124} {\bibfield
  {journal} {\bibinfo  {journal} {Phys. Rev. B}\ }\textbf {\bibinfo {volume}
  {85}},\ \bibinfo {pages} {045124} (\bibinfo {year} {2012})}\BibitemShut
  {NoStop}%
\bibitem [{\citenamefont {Xu}\ \emph {et~al.}(2015)\citenamefont {Xu},
  \citenamefont {Alidoust}, \citenamefont {Belopolski}, \citenamefont {Yuan},
  \citenamefont {Bian}, \citenamefont {Chang}, \citenamefont {Zheng},
  \citenamefont {Strocov}, \citenamefont {Sanchez}, \citenamefont {Chang},
  \citenamefont {Zhang}, \citenamefont {Mou}, \citenamefont {Wu}, \citenamefont
  {Huang}, \citenamefont {Lee}, \citenamefont {Huang}, \citenamefont {Wang},
  \citenamefont {Bansil}, \citenamefont {Jeng}, \citenamefont {Neupert},
  \citenamefont {Kaminski}, \citenamefont {Lin}, \citenamefont {Jia},\ and\
  \citenamefont {Hasan}}]{Xu2015}%
  \BibitemOpen
  \bibfield  {author} {\bibinfo {author} {\bibfnamefont {Su-Yang}\ \bibnamefont
  {Xu}}, \bibinfo {author} {\bibfnamefont {Nasser}\ \bibnamefont {Alidoust}},
  \bibinfo {author} {\bibfnamefont {Ilya}\ \bibnamefont {Belopolski}}, \bibinfo
  {author} {\bibfnamefont {Zhujun}\ \bibnamefont {Yuan}}, \bibinfo {author}
  {\bibfnamefont {Guang}\ \bibnamefont {Bian}}, \bibinfo {author}
  {\bibfnamefont {Tay-Rong}\ \bibnamefont {Chang}}, \bibinfo {author}
  {\bibfnamefont {Hao}\ \bibnamefont {Zheng}}, \bibinfo {author} {\bibfnamefont
  {Vladimir~N}\ \bibnamefont {Strocov}}, \bibinfo {author} {\bibfnamefont
  {Daniel~S}\ \bibnamefont {Sanchez}}, \bibinfo {author} {\bibfnamefont
  {Guoqing}\ \bibnamefont {Chang}}, \bibinfo {author} {\bibfnamefont
  {Chenglong}\ \bibnamefont {Zhang}}, \bibinfo {author} {\bibfnamefont
  {Daixiang}\ \bibnamefont {Mou}}, \bibinfo {author} {\bibfnamefont {Yun}\
  \bibnamefont {Wu}}, \bibinfo {author} {\bibfnamefont {Lunan}\ \bibnamefont
  {Huang}}, \bibinfo {author} {\bibfnamefont {Chi-Cheng}\ \bibnamefont {Lee}},
  \bibinfo {author} {\bibfnamefont {Shin-Ming}\ \bibnamefont {Huang}}, \bibinfo
  {author} {\bibfnamefont {BaoKai}\ \bibnamefont {Wang}}, \bibinfo {author}
  {\bibfnamefont {Arun}\ \bibnamefont {Bansil}}, \bibinfo {author}
  {\bibfnamefont {Horng-Tay}\ \bibnamefont {Jeng}}, \bibinfo {author}
  {\bibfnamefont {Titus}\ \bibnamefont {Neupert}}, \bibinfo {author}
  {\bibfnamefont {Adam}\ \bibnamefont {Kaminski}}, \bibinfo {author}
  {\bibfnamefont {Hsin}\ \bibnamefont {Lin}}, \bibinfo {author} {\bibfnamefont
  {Shuang}\ \bibnamefont {Jia}}, \ and\ \bibinfo {author} {\bibfnamefont
  {Zahid~M.}\ \bibnamefont {Hasan}},\ }\bibfield  {title} {\enquote {\bibinfo
  {title} {Discovery of a weyl fermion state with fermi arcs in niobium
  arsenide},}\ }\href {\doibase 10.1038/nphys3437} {\bibfield  {journal}
  {\bibinfo  {journal} {Nature Phys.}\ }\textbf {\bibinfo {volume} {11}},\
  \bibinfo {pages} {748} (\bibinfo {year} {2015})}\BibitemShut {NoStop}%
\bibitem [{\citenamefont {Lv}\ \emph {et~al.}(2015)\citenamefont {Lv},
  \citenamefont {Weng}, \citenamefont {Fu}, \citenamefont {Wang}, \citenamefont
  {Miao}, \citenamefont {Ma}, \citenamefont {Richard}, \citenamefont {Huang},
  \citenamefont {Zhao}, \citenamefont {Chen}, \citenamefont {Fang},
  \citenamefont {Dai}, \citenamefont {Qian},\ and\ \citenamefont
  {Ding}}]{Lv2015}%
  \BibitemOpen
  \bibfield  {author} {\bibinfo {author} {\bibfnamefont {B.~Q.}\ \bibnamefont
  {Lv}}, \bibinfo {author} {\bibfnamefont {H.~M.}\ \bibnamefont {Weng}},
  \bibinfo {author} {\bibfnamefont {B.~B.}\ \bibnamefont {Fu}}, \bibinfo
  {author} {\bibfnamefont {X.~P.}\ \bibnamefont {Wang}}, \bibinfo {author}
  {\bibfnamefont {H.}~\bibnamefont {Miao}}, \bibinfo {author} {\bibfnamefont
  {J.}~\bibnamefont {Ma}}, \bibinfo {author} {\bibfnamefont {P.}~\bibnamefont
  {Richard}}, \bibinfo {author} {\bibfnamefont {X.~C.}\ \bibnamefont {Huang}},
  \bibinfo {author} {\bibfnamefont {L.~X.}\ \bibnamefont {Zhao}}, \bibinfo
  {author} {\bibfnamefont {G.~F.}\ \bibnamefont {Chen}}, \bibinfo {author}
  {\bibfnamefont {Z.}~\bibnamefont {Fang}}, \bibinfo {author} {\bibfnamefont
  {X.}~\bibnamefont {Dai}}, \bibinfo {author} {\bibfnamefont {T.}~\bibnamefont
  {Qian}}, \ and\ \bibinfo {author} {\bibfnamefont {H.}~\bibnamefont {Ding}},\
  }\bibfield  {title} {\enquote {\bibinfo {title} {Experimental discovery of
  weyl semimetal {TaAs}},}\ }\href {\doibase 10.1103/PhysRevX.5.031013}
  {\bibfield  {journal} {\bibinfo  {journal} {Phys. Rev. X}\ }\textbf {\bibinfo
  {volume} {5}},\ \bibinfo {pages} {031013} (\bibinfo {year}
  {2015})}\BibitemShut {NoStop}%
\bibitem [{\citenamefont {Lu}\ \emph {et~al.}(2015)\citenamefont {Lu},
  \citenamefont {Wang}, \citenamefont {Ye}, \citenamefont {Ran}, \citenamefont
  {Fu}, \citenamefont {Joannopoulos},\ and\ \citenamefont {Solja{\v
  c}i{\'c}}}]{Lu2015a}%
  \BibitemOpen
  \bibfield  {author} {\bibinfo {author} {\bibfnamefont {Ling}\ \bibnamefont
  {Lu}}, \bibinfo {author} {\bibfnamefont {Zhiyu}\ \bibnamefont {Wang}},
  \bibinfo {author} {\bibfnamefont {Dexin}\ \bibnamefont {Ye}}, \bibinfo
  {author} {\bibfnamefont {Lixin}\ \bibnamefont {Ran}}, \bibinfo {author}
  {\bibfnamefont {Liang}\ \bibnamefont {Fu}}, \bibinfo {author} {\bibfnamefont
  {John~D.}\ \bibnamefont {Joannopoulos}}, \ and\ \bibinfo {author}
  {\bibfnamefont {Marin}\ \bibnamefont {Solja{\v c}i{\'c}}},\ }\bibfield
  {title} {\enquote {\bibinfo {title} {{Experimental observation of Weyl
  points}},}\ }\href {\doibase 10.1126/science.aaa9273} {\bibfield  {journal}
  {\bibinfo  {journal} {Science}\ }\textbf {\bibinfo {volume} {349}},\ \bibinfo
  {pages} {622--624} (\bibinfo {year} {2015})}\BibitemShut {NoStop}%
\bibitem [{\citenamefont {Bradlyn}\ \emph {et~al.}(2017)\citenamefont
  {Bradlyn}, \citenamefont {Elcoro}, \citenamefont {Cano}, \citenamefont
  {Vergniory}, \citenamefont {Wang}, \citenamefont {Felser}, \citenamefont
  {Aroyo},\ and\ \citenamefont {Bernevig}}]{Bradlyn2017}%
  \BibitemOpen
  \bibfield  {author} {\bibinfo {author} {\bibfnamefont {Barry}\ \bibnamefont
  {Bradlyn}}, \bibinfo {author} {\bibfnamefont {L}~\bibnamefont {Elcoro}},
  \bibinfo {author} {\bibfnamefont {Jennifer}\ \bibnamefont {Cano}}, \bibinfo
  {author} {\bibfnamefont {MG}~\bibnamefont {Vergniory}}, \bibinfo {author}
  {\bibfnamefont {Zhijun}\ \bibnamefont {Wang}}, \bibinfo {author}
  {\bibfnamefont {C}~\bibnamefont {Felser}}, \bibinfo {author} {\bibfnamefont
  {MI}~\bibnamefont {Aroyo}}, \ and\ \bibinfo {author} {\bibfnamefont
  {B~Andrei}\ \bibnamefont {Bernevig}},\ }\bibfield  {title} {\enquote
  {\bibinfo {title} {Topological quantum chemistry},}\ }\href {\doibase
  https://doi.org/10.1038/nature23268} {\bibfield  {journal} {\bibinfo
  {journal} {Nature}\ }\textbf {\bibinfo {volume} {547}},\ \bibinfo {pages}
  {298} (\bibinfo {year} {2017})}\BibitemShut {NoStop}%
\bibitem [{\citenamefont {Armitage}\ \emph {et~al.}(2018)\citenamefont
  {Armitage}, \citenamefont {Mele},\ and\ \citenamefont
  {Vishwanath}}]{Armitage2018}%
  \BibitemOpen
  \bibfield  {author} {\bibinfo {author} {\bibfnamefont {N.~P.}\ \bibnamefont
  {Armitage}}, \bibinfo {author} {\bibfnamefont {E.~J.}\ \bibnamefont {Mele}},
  \ and\ \bibinfo {author} {\bibfnamefont {Ashvin}\ \bibnamefont
  {Vishwanath}},\ }\bibfield  {title} {\enquote {\bibinfo {title} {Weyl and
  dirac semimetals in three-dimensional solids},}\ }\href {\doibase
  10.1103/RevModPhys.90.015001} {\bibfield  {journal} {\bibinfo  {journal}
  {Rev. Mod. Phys.}\ }\textbf {\bibinfo {volume} {90}},\ \bibinfo {pages}
  {015001} (\bibinfo {year} {2018})}\BibitemShut {NoStop}%
\bibitem [{\citenamefont {Li}\ and\ \citenamefont {Haldane}(2018)}]{Li2018}%
  \BibitemOpen
  \bibfield  {author} {\bibinfo {author} {\bibfnamefont {Yi}~\bibnamefont
  {Li}}\ and\ \bibinfo {author} {\bibfnamefont {F.~D.~M.}\ \bibnamefont
  {Haldane}},\ }\bibfield  {title} {\enquote {\bibinfo {title} {Topological
  nodal cooper pairing in doped weyl metals},}\ }\href {\doibase
  10.1103/PhysRevLett.120.067003} {\bibfield  {journal} {\bibinfo  {journal}
  {Phys. Rev. Lett.}\ }\textbf {\bibinfo {volume} {120}},\ \bibinfo {pages}
  {067003} (\bibinfo {year} {2018})}\BibitemShut {NoStop}%
\bibitem [{\citenamefont {Sun}\ \emph {et~al.}(2019)\citenamefont {Sun},
  \citenamefont {Lee},\ and\ \citenamefont {Li}}]{Sun2019}%
  \BibitemOpen
  \bibfield  {author} {\bibinfo {author} {\bibfnamefont {Canon}\ \bibnamefont
  {Sun}}, \bibinfo {author} {\bibfnamefont {Shu-Ping}\ \bibnamefont {Lee}}, \
  and\ \bibinfo {author} {\bibfnamefont {Yi}~\bibnamefont {Li}},\ }\bibfield
  {title} {\enquote {\bibinfo {title} {Vortices in a monopole superconducting
  weyl semi-metal},}\ }\href@noop {} {\  (\bibinfo {year} {2019})},\ \Eprint
  {http://arxiv.org/abs/1909.04179v2} {1909.04179v2} \BibitemShut {NoStop}%
\bibitem [{\citenamefont {Bobrow}\ and\ \citenamefont {Li}(2022)}]{Bobrow2022}%
  \BibitemOpen
  \bibfield  {author} {\bibinfo {author} {\bibfnamefont {Eric}\ \bibnamefont
  {Bobrow}}\ and\ \bibinfo {author} {\bibfnamefont {Yi}~\bibnamefont {Li}},\
  }\bibfield  {title} {\enquote {\bibinfo {title} {Monopole superconductivity
  in magnetically doped {Cd$_3$As$_2$}},}\ }\href@noop {} {\  (\bibinfo {year}
  {2022})},\ \Eprint {http://arxiv.org/abs/2204.04249v1} {2204.04249v1}
  \BibitemShut {NoStop}%
\bibitem [{\citenamefont {Frazier}\ \emph {et~al.}(2024)\citenamefont
  {Frazier}, \citenamefont {Zhang}, \citenamefont {Zhang}, \citenamefont
  {Sun},\ and\ \citenamefont {Li}}]{Frazier2024}%
  \BibitemOpen
  \bibfield  {author} {\bibinfo {author} {\bibfnamefont {Grayson~R.}\
  \bibnamefont {Frazier}}, \bibinfo {author} {\bibfnamefont {Junjia}\
  \bibnamefont {Zhang}}, \bibinfo {author} {\bibfnamefont {Junyi}\ \bibnamefont
  {Zhang}}, \bibinfo {author} {\bibfnamefont {Xinyu}\ \bibnamefont {Sun}}, \
  and\ \bibinfo {author} {\bibfnamefont {Yi}~\bibnamefont {Li}},\ }\bibfield
  {title} {\enquote {\bibinfo {title} {Designing phase sensitive probes of
  monopole superconducting order},}\ }\href {\doibase
  10.1103/physrevresearch.6.043189} {\bibfield  {journal} {\bibinfo  {journal}
  {Physical Review Research}\ }\textbf {\bibinfo {volume} {6}} (\bibinfo {year}
  {2024}),\ 10.1103/physrevresearch.6.043189}\BibitemShut {NoStop}%
\bibitem [{\citenamefont {Murakami}\ and\ \citenamefont
  {Nagaosa}(2003)}]{Murakami2003a}%
  \BibitemOpen
  \bibfield  {author} {\bibinfo {author} {\bibfnamefont {Shuichi}\ \bibnamefont
  {Murakami}}\ and\ \bibinfo {author} {\bibfnamefont {Naoto}\ \bibnamefont
  {Nagaosa}},\ }\bibfield  {title} {\enquote {\bibinfo {title} {Berry phase in
  magnetic superconductors},}\ }\href {\doibase 10.1103/PhysRevLett.90.057002}
  {\bibfield  {journal} {\bibinfo  {journal} {Phys. Rev. Lett.}\ }\textbf
  {\bibinfo {volume} {90}},\ \bibinfo {pages} {057002} (\bibinfo {year}
  {2003})}\BibitemShut {NoStop}%
\bibitem [{\citenamefont {Dirac}(1931)}]{Dirac1931}%
  \BibitemOpen
  \bibfield  {author} {\bibinfo {author} {\bibfnamefont {Paul Adrien~Maurice}\
  \bibnamefont {Dirac}},\ }\bibfield  {title} {\enquote {\bibinfo {title}
  {Quantised singularities in the electromagnetic field},}\ }\href {\doibase
  10.1098/rspa.1931.0130} {\bibfield  {journal} {\bibinfo  {journal} {Proc. R.
  Soc. Lond. A.}\ }\textbf {\bibinfo {volume} {133}},\ \bibinfo {pages}
  {60--72} (\bibinfo {year} {1931})}\BibitemShut {NoStop}%
\bibitem [{\citenamefont {Tamm}(1931)}]{Tamm1931}%
  \BibitemOpen
  \bibfield  {author} {\bibinfo {author} {\bibfnamefont {Ig}~\bibnamefont
  {Tamm}},\ }\bibfield  {title} {\enquote {\bibinfo {title} {{Die
  verallgemeinerten Kugelfunktionen und die Wellenfunktionen eines Elektrons im
  Felde eines Magnetpoles}},}\ }\href {\doibase 10.1007/bf01341701} {\bibfield
  {journal} {\bibinfo  {journal} {Zeitschrift F\"ur Physik}\ }\textbf {\bibinfo
  {volume} {71}},\ \bibinfo {pages} {141--150} (\bibinfo {year}
  {1931})}\BibitemShut {NoStop}%
\bibitem [{\citenamefont {Wu}\ and\ \citenamefont {Yang}(1976)}]{Wu1976}%
  \BibitemOpen
  \bibfield  {author} {\bibinfo {author} {\bibfnamefont {Tai~Tsun}\
  \bibnamefont {Wu}}\ and\ \bibinfo {author} {\bibfnamefont {Chen~Ning}\
  \bibnamefont {Yang}},\ }\bibfield  {title} {\enquote {\bibinfo {title} {Dirac
  monopole without strings: Monopole harmonics},}\ }\href {\doibase
  http://dx.doi.org/10.1016/0550-3213(76)90143-7} {\bibfield  {journal}
  {\bibinfo  {journal} {Nucl. Phys. B}\ }\textbf {\bibinfo {volume} {107}},\
  \bibinfo {pages} {365 -- 380} (\bibinfo {year} {1976})}\BibitemShut {NoStop}%
\bibitem [{\citenamefont {Haldane}(1983)}]{Haldane1983}%
  \BibitemOpen
  \bibfield  {author} {\bibinfo {author} {\bibfnamefont {F.~D.~M.}\
  \bibnamefont {Haldane}},\ }\bibfield  {title} {\enquote {\bibinfo {title}
  {Fractional quantization of the hall effect: A hierarchy of incompressible
  quantum fluid states},}\ }\href {\doibase 10.1103/PhysRevLett.51.605}
  {\bibfield  {journal} {\bibinfo  {journal} {Phys. Rev. Lett.}\ }\textbf
  {\bibinfo {volume} {51}},\ \bibinfo {pages} {605--608} (\bibinfo {year}
  {1983})}\BibitemShut {NoStop}%
\bibitem [{\citenamefont {Bobrow}\ \emph {et~al.}(2020)\citenamefont {Bobrow},
  \citenamefont {Sun},\ and\ \citenamefont {Li}}]{Bobrow2020}%
  \BibitemOpen
  \bibfield  {author} {\bibinfo {author} {\bibfnamefont {Eric}\ \bibnamefont
  {Bobrow}}, \bibinfo {author} {\bibfnamefont {Canon}\ \bibnamefont {Sun}}, \
  and\ \bibinfo {author} {\bibfnamefont {Yi}~\bibnamefont {Li}},\ }\bibfield
  {title} {\enquote {\bibinfo {title} {Monopole charge density wave states in
  weyl semimetals},}\ }\href {\doibase 10.1103/PhysRevResearch.2.012078}
  {\bibfield  {journal} {\bibinfo  {journal} {Phys. Rev. Res.}\ }\textbf
  {\bibinfo {volume} {2}},\ \bibinfo {pages} {012078} (\bibinfo {year}
  {2020})}\BibitemShut {NoStop}%
\bibitem [{\citenamefont {Read}\ and\ \citenamefont {Green}(2000)}]{Read2000}%
  \BibitemOpen
  \bibfield  {author} {\bibinfo {author} {\bibfnamefont {N.}~\bibnamefont
  {Read}}\ and\ \bibinfo {author} {\bibfnamefont {D.}~\bibnamefont {Green}},\
  }\bibfield  {title} {\enquote {\bibinfo {title} {{Paired states of fermions
  in two dimensions with breaking of parity and time-reversal symmetries and
  the fractional quantum Hall effect}},}\ }\href
  {https://journals.aps.org/prb/abstract/10.1103/PhysRevB.61.10267} {\bibfield
  {journal} {\bibinfo  {journal} {Phys. Rev. B}\ }\textbf {\bibinfo {volume}
  {61}},\ \bibinfo {pages} {10267--10297} (\bibinfo {year} {2000})}\BibitemShut
  {NoStop}%
\bibitem [{\citenamefont {Fu}\ and\ \citenamefont {Kane}(2008)}]{Fu2008}%
  \BibitemOpen
  \bibfield  {author} {\bibinfo {author} {\bibfnamefont {L.}~\bibnamefont
  {Fu}}\ and\ \bibinfo {author} {\bibfnamefont {CL}~\bibnamefont {Kane}},\
  }\bibfield  {title} {\enquote {\bibinfo {title} {{Superconducting proximity
  effect and Majorana fermions at the surface of a topological insulator}},}\
  }\href {https://journals.aps.org/prl/abstract/10.1103/PhysRevLett.100.096407}
  {\bibfield  {journal} {\bibinfo  {journal} {Phys. Rev. Lett.}\ }\textbf
  {\bibinfo {volume} {100}},\ \bibinfo {pages} {96407} (\bibinfo {year}
  {2008})}\BibitemShut {NoStop}%
\bibitem [{\citenamefont {Schnyder}\ \emph {et~al.}(2008)\citenamefont
  {Schnyder}, \citenamefont {Ryu}, \citenamefont {Furusaki},\ and\
  \citenamefont {Ludwig}}]{Schnyder2008}%
  \BibitemOpen
  \bibfield  {author} {\bibinfo {author} {\bibfnamefont {A.P.}\ \bibnamefont
  {Schnyder}}, \bibinfo {author} {\bibfnamefont {S.}~\bibnamefont {Ryu}},
  \bibinfo {author} {\bibfnamefont {A.}~\bibnamefont {Furusaki}}, \ and\
  \bibinfo {author} {\bibfnamefont {A.W.W.}\ \bibnamefont {Ludwig}},\
  }\bibfield  {title} {\enquote {\bibinfo {title} {{Classification of
  topological insulators and superconductors in three spatial dimensions}},}\
  }\href@noop {} {\bibfield  {journal} {\bibinfo  {journal} {Phys. Rev. B}\
  }\textbf {\bibinfo {volume} {78}},\ \bibinfo {pages} {195125} (\bibinfo
  {year} {2008})}\BibitemShut {NoStop}%
\bibitem [{\citenamefont {Chung}\ and\ \citenamefont
  {Zhang}(2009)}]{Chung2009}%
  \BibitemOpen
  \bibfield  {author} {\bibinfo {author} {\bibfnamefont {Suk~Bum}\ \bibnamefont
  {Chung}}\ and\ \bibinfo {author} {\bibfnamefont {Shou-Cheng}\ \bibnamefont
  {Zhang}},\ }\bibfield  {title} {\enquote {\bibinfo {title} {Detecting the
  majorana fermion surface state of he 3- b through spin relaxation},}\ }\href
  {https://journals.aps.org/prl/abstract/10.1103/PhysRevLett.103.235301}
  {\bibfield  {journal} {\bibinfo  {journal} {Phys. Rev. Lett.}\ }\textbf
  {\bibinfo {volume} {103}},\ \bibinfo {pages} {235301} (\bibinfo {year}
  {2009})}\BibitemShut {NoStop}%
\bibitem [{\citenamefont {Lutchyn}\ \emph {et~al.}(2010)\citenamefont
  {Lutchyn}, \citenamefont {Sau},\ and\ \citenamefont {Sarma}}]{Lutchyn2010}%
  \BibitemOpen
  \bibfield  {author} {\bibinfo {author} {\bibfnamefont {Roman~M}\ \bibnamefont
  {Lutchyn}}, \bibinfo {author} {\bibfnamefont {Jay~D}\ \bibnamefont {Sau}}, \
  and\ \bibinfo {author} {\bibfnamefont {S~Das}\ \bibnamefont {Sarma}},\
  }\bibfield  {title} {\enquote {\bibinfo {title} {Majorana fermions and a
  topological phase transition in semiconductor-superconductor
  heterostructures},}\ }\href
  {https://journals.aps.org/prl/pdf/10.1103/PhysRevLett.105.077001} {\bibfield
  {journal} {\bibinfo  {journal} {Phys. Rev. Lett.}\ }\textbf {\bibinfo
  {volume} {105}},\ \bibinfo {pages} {077001} (\bibinfo {year}
  {2010})}\BibitemShut {NoStop}%
\bibitem [{\citenamefont {Qi}\ and\ \citenamefont {Zhang}(2011)}]{Qi2011}%
  \BibitemOpen
  \bibfield  {author} {\bibinfo {author} {\bibfnamefont {Xiao-Liang}\
  \bibnamefont {Qi}}\ and\ \bibinfo {author} {\bibfnamefont {Shou-Cheng}\
  \bibnamefont {Zhang}},\ }\bibfield  {title} {\enquote {\bibinfo {title}
  {Topological insulators and superconductors},}\ }\href {\doibase
  10.1103/RevModPhys.83.1057} {\bibfield  {journal} {\bibinfo  {journal} {Rev.
  Mod. Phys.}\ }\textbf {\bibinfo {volume} {83}},\ \bibinfo {pages}
  {1057--1110} (\bibinfo {year} {2011})}\BibitemShut {NoStop}%
\bibitem [{\citenamefont {Alicea}(2012)}]{Alicea2012}%
  \BibitemOpen
  \bibfield  {author} {\bibinfo {author} {\bibfnamefont {Jason}\ \bibnamefont
  {Alicea}},\ }\bibfield  {title} {\enquote {\bibinfo {title} {New directions
  in the pursuit of majorana fermions in solid state systems},}\ }\href
  {https://iopscience.iop.org/article/10.1088/0034-4885/75/7/076501/meta}
  {\bibfield  {journal} {\bibinfo  {journal} {Reports on Progress in Physics}\
  }\textbf {\bibinfo {volume} {75}},\ \bibinfo {pages} {076501} (\bibinfo
  {year} {2012})}\BibitemShut {NoStop}%
\bibitem [{\citenamefont {Sato}\ and\ \citenamefont {Ando}(2017)}]{Sato2017}%
  \BibitemOpen
  \bibfield  {author} {\bibinfo {author} {\bibfnamefont {Masatoshi}\
  \bibnamefont {Sato}}\ and\ \bibinfo {author} {\bibfnamefont {Yoichi}\
  \bibnamefont {Ando}},\ }\bibfield  {title} {\enquote {\bibinfo {title}
  {Topological superconductors: a review},}\ }\href
  {https://iopscience.iop.org/article/10.1088/1361-6633/aa6ac7/meta} {\bibfield
   {journal} {\bibinfo  {journal} {Reports on Progress in Physics}\ }\textbf
  {\bibinfo {volume} {80}},\ \bibinfo {pages} {076501} (\bibinfo {year}
  {2017})}\BibitemShut {NoStop}%
\bibitem [{\citenamefont {Nielsen}\ \emph
  {et~al.}(1981{\natexlab{a}})\citenamefont {Nielsen} \emph
  {et~al.}}]{Nielsen1981a}%
  \BibitemOpen
  \bibfield  {author} {\bibinfo {author} {\bibfnamefont {M.}~\bibnamefont
  {Nielsen}} \emph {et~al.},\ }\bibfield  {title} {\enquote {\bibinfo {title}
  {{Absence of neutrinos on a lattice::(I). Proof by homotopy theory}},}\
  }\href {\doibase https://doi.org/10.1016/0550-3213(81)90361-8} {\bibfield
  {journal} {\bibinfo  {journal} {Nucl. Phys. B}\ }\textbf {\bibinfo {volume}
  {185}},\ \bibinfo {pages} {20--40} (\bibinfo {year}
  {1981}{\natexlab{a}})}\BibitemShut {NoStop}%
\bibitem [{\citenamefont {Nielsen}\ \emph
  {et~al.}(1981{\natexlab{b}})\citenamefont {Nielsen} \emph
  {et~al.}}]{Nielsen1981b}%
  \BibitemOpen
  \bibfield  {author} {\bibinfo {author} {\bibfnamefont {M.}~\bibnamefont
  {Nielsen}} \emph {et~al.},\ }\bibfield  {title} {\enquote {\bibinfo {title}
  {{Absence of neutrinos on a lattice::(II). Intuitive topological proof}},}\
  }\href {\doibase https://doi.org/10.1016/0550-3213(81)90524-1} {\bibfield
  {journal} {\bibinfo  {journal} {Nucl. Phys. B}\ }\textbf {\bibinfo {volume}
  {193}},\ \bibinfo {pages} {173--194} (\bibinfo {year}
  {1981}{\natexlab{b}})}\BibitemShut {NoStop}%
\bibitem [{\citenamefont {Mermin}\ and\ \citenamefont {Ho}(1976)}]{Mermin1976}%
  \BibitemOpen
  \bibfield  {author} {\bibinfo {author} {\bibfnamefont {N.~D.}\ \bibnamefont
  {Mermin}}\ and\ \bibinfo {author} {\bibfnamefont {Tin-Lun}\ \bibnamefont
  {Ho}},\ }\bibfield  {title} {\enquote {\bibinfo {title} {Circulation and
  angular momentum in the {$A$} phase of superfluid helium-3},}\ }\href
  {\doibase 10.1103/PhysRevLett.36.594} {\bibfield  {journal} {\bibinfo
  {journal} {Phys. Rev. Lett.}\ }\textbf {\bibinfo {volume} {36}},\ \bibinfo
  {pages} {594--597} (\bibinfo {year} {1976})}\BibitemShut {NoStop}%
\bibitem [{\citenamefont {Lin}\ \emph {et~al.}(2011)\citenamefont {Lin},
  \citenamefont {Jim{\'e}nez-Garc{\'\i}a},\ and\ \citenamefont
  {Spielman}}]{Lin2011}%
  \BibitemOpen
  \bibfield  {author} {\bibinfo {author} {\bibfnamefont {Y.~J.}\ \bibnamefont
  {Lin}}, \bibinfo {author} {\bibfnamefont {K.}~\bibnamefont
  {Jim{\'e}nez-Garc{\'\i}a}}, \ and\ \bibinfo {author} {\bibfnamefont {I.~B.}\
  \bibnamefont {Spielman}},\ }\bibfield  {title} {\enquote {\bibinfo {title}
  {Spin-orbit-coupled bose-einstein condensates},}\ }\href@noop {} {\bibfield
  {journal} {\bibinfo  {journal} {Nature}\ }\textbf {\bibinfo {volume} {471}},\
  \bibinfo {pages} {83--86} (\bibinfo {year} {2011})}\BibitemShut {NoStop}%
\bibitem [{\citenamefont {Li}\ \emph {et~al.}(2012)\citenamefont {Li},
  \citenamefont {Intriligator}, \citenamefont {Yu},\ and\ \citenamefont
  {Wu}}]{Li2012}%
  \BibitemOpen
  \bibfield  {author} {\bibinfo {author} {\bibfnamefont {Yi}~\bibnamefont
  {Li}}, \bibinfo {author} {\bibfnamefont {Kenneth}\ \bibnamefont
  {Intriligator}}, \bibinfo {author} {\bibfnamefont {Yue}\ \bibnamefont {Yu}},
  \ and\ \bibinfo {author} {\bibfnamefont {Congjun}\ \bibnamefont {Wu}},\
  }\bibfield  {title} {\enquote {\bibinfo {title} {Isotropic landau levels of
  dirac fermions in high dimensions},}\ }\href {\doibase
  10.1103/PhysRevB.85.085132} {\bibfield  {journal} {\bibinfo  {journal} {Phys.
  Rev. B}\ }\textbf {\bibinfo {volume} {85}},\ \bibinfo {pages} {085132}
  (\bibinfo {year} {2012})}\BibitemShut {NoStop}%
\bibitem [{\citenamefont {Anderson}\ \emph {et~al.}(2012)\citenamefont
  {Anderson}, \citenamefont {Juzeli\ifmmode~\bar{u}\else \={u}\fi{}nas},
  \citenamefont {Galitski},\ and\ \citenamefont {Spielman}}]{Anderson2012}%
  \BibitemOpen
  \bibfield  {author} {\bibinfo {author} {\bibfnamefont {Brandon~M.}\
  \bibnamefont {Anderson}}, \bibinfo {author} {\bibfnamefont {Gediminas}\
  \bibnamefont {Juzeli\ifmmode~\bar{u}\else \={u}\fi{}nas}}, \bibinfo {author}
  {\bibfnamefont {Victor~M.}\ \bibnamefont {Galitski}}, \ and\ \bibinfo
  {author} {\bibfnamefont {I.~B.}\ \bibnamefont {Spielman}},\ }\bibfield
  {title} {\enquote {\bibinfo {title} {Synthetic 3d spin-orbit coupling},}\
  }\href {\doibase 10.1103/PhysRevLett.108.235301} {\bibfield  {journal}
  {\bibinfo  {journal} {Phys. Rev. Lett.}\ }\textbf {\bibinfo {volume} {108}},\
  \bibinfo {pages} {235301} (\bibinfo {year} {2012})}\BibitemShut {NoStop}%
\bibitem [{\citenamefont {Wang}\ \emph {et~al.}(2012)\citenamefont {Wang},
  \citenamefont {Yu}, \citenamefont {Fu}, \citenamefont {Miao}, \citenamefont
  {Huang}, \citenamefont {Chai}, \citenamefont {Zhai},\ and\ \citenamefont
  {Zhang}}]{Wang2012a}%
  \BibitemOpen
  \bibfield  {author} {\bibinfo {author} {\bibfnamefont {Pengjun}\ \bibnamefont
  {Wang}}, \bibinfo {author} {\bibfnamefont {Zeng-Qiang}\ \bibnamefont {Yu}},
  \bibinfo {author} {\bibfnamefont {Zhengkun}\ \bibnamefont {Fu}}, \bibinfo
  {author} {\bibfnamefont {Jiao}\ \bibnamefont {Miao}}, \bibinfo {author}
  {\bibfnamefont {Lianghui}\ \bibnamefont {Huang}}, \bibinfo {author}
  {\bibfnamefont {Shijie}\ \bibnamefont {Chai}}, \bibinfo {author}
  {\bibfnamefont {Hui}\ \bibnamefont {Zhai}}, \ and\ \bibinfo {author}
  {\bibfnamefont {Jing}\ \bibnamefont {Zhang}},\ }\bibfield  {title} {\enquote
  {\bibinfo {title} {Spin-orbit coupled degenerate fermi gases},}\ }\href
  {\doibase 10.1103/physrevlett.109.095301} {\bibfield  {journal} {\bibinfo
  {journal} {Phys. Rev. Lett. 109, 095301 (2012)}\ }\textbf {\bibinfo {volume}
  {109}},\ \bibinfo {pages} {095301} (\bibinfo {year} {2012})}\BibitemShut
  {NoStop}%
\bibitem [{\citenamefont {Cheuk}\ \emph {et~al.}(2012)\citenamefont {Cheuk},
  \citenamefont {Sommer}, \citenamefont {Hadzibabic}, \citenamefont {Yefsah},
  \citenamefont {Bakr},\ and\ \citenamefont {Zwierlein}}]{Cheuk2012}%
  \BibitemOpen
  \bibfield  {author} {\bibinfo {author} {\bibfnamefont {Lawrence~W.}\
  \bibnamefont {Cheuk}}, \bibinfo {author} {\bibfnamefont {Ariel~T.}\
  \bibnamefont {Sommer}}, \bibinfo {author} {\bibfnamefont {Zoran}\
  \bibnamefont {Hadzibabic}}, \bibinfo {author} {\bibfnamefont {Tarik}\
  \bibnamefont {Yefsah}}, \bibinfo {author} {\bibfnamefont {Waseem~S.}\
  \bibnamefont {Bakr}}, \ and\ \bibinfo {author} {\bibfnamefont {Martin~W.}\
  \bibnamefont {Zwierlein}},\ }\bibfield  {title} {\enquote {\bibinfo {title}
  {Spin-injection spectroscopy of a spin-orbit coupled fermi gas},}\ }\href
  {\doibase 10.1103/physrevlett.109.095302} {\bibfield  {journal} {\bibinfo
  {journal} {Physical Review Letters}\ }\textbf {\bibinfo {volume} {109}},\
  \bibinfo {pages} {095302} (\bibinfo {year} {2012})}\BibitemShut {NoStop}%
\bibitem [{\citenamefont {Song}\ \emph {et~al.}(2019)\citenamefont {Song},
  \citenamefont {He}, \citenamefont {Niu}, \citenamefont {Zhang}, \citenamefont
  {Ren}, \citenamefont {Liu},\ and\ \citenamefont {Jo}}]{Song2019}%
  \BibitemOpen
  \bibfield  {author} {\bibinfo {author} {\bibfnamefont {Bo}~\bibnamefont
  {Song}}, \bibinfo {author} {\bibfnamefont {Chengdong}\ \bibnamefont {He}},
  \bibinfo {author} {\bibfnamefont {Sen}\ \bibnamefont {Niu}}, \bibinfo
  {author} {\bibfnamefont {Long}\ \bibnamefont {Zhang}}, \bibinfo {author}
  {\bibfnamefont {Zejian}\ \bibnamefont {Ren}}, \bibinfo {author}
  {\bibfnamefont {Xiong-Jun}\ \bibnamefont {Liu}}, \ and\ \bibinfo {author}
  {\bibfnamefont {Gyu-Boong}\ \bibnamefont {Jo}},\ }\bibfield  {title}
  {\enquote {\bibinfo {title} {Observation of nodal-line semimetal with
  ultracold fermions in an optical lattice},}\ }\href {\doibase
  10.1038/s41567-019-0564-y} {\bibfield  {journal} {\bibinfo  {journal} {Nature
  Physics}\ }\textbf {\bibinfo {volume} {15}},\ \bibinfo {pages} {911--916}
  (\bibinfo {year} {2019})}\BibitemShut {NoStop}%
\bibitem [{\citenamefont {Wang}\ \emph {et~al.}(2021)\citenamefont {Wang},
  \citenamefont {Cheng}, \citenamefont {Wang}, \citenamefont {Zhang},
  \citenamefont {Lu}, \citenamefont {Yi}, \citenamefont {Niu}, \citenamefont
  {Deng}, \citenamefont {Liu}, \citenamefont {Chen},\ and\ \citenamefont
  {Pan}}]{Wang2021}%
  \BibitemOpen
  \bibfield  {author} {\bibinfo {author} {\bibfnamefont {Zong-Yao}\
  \bibnamefont {Wang}}, \bibinfo {author} {\bibfnamefont {Xiang-Can}\
  \bibnamefont {Cheng}}, \bibinfo {author} {\bibfnamefont {Bao-Zong}\
  \bibnamefont {Wang}}, \bibinfo {author} {\bibfnamefont {Jin-Yi}\ \bibnamefont
  {Zhang}}, \bibinfo {author} {\bibfnamefont {Yue-Hui}\ \bibnamefont {Lu}},
  \bibinfo {author} {\bibfnamefont {Chang-Rui}\ \bibnamefont {Yi}}, \bibinfo
  {author} {\bibfnamefont {Sen}\ \bibnamefont {Niu}}, \bibinfo {author}
  {\bibfnamefont {Youjin}\ \bibnamefont {Deng}}, \bibinfo {author}
  {\bibfnamefont {Xiong-Jun}\ \bibnamefont {Liu}}, \bibinfo {author}
  {\bibfnamefont {Shuai}\ \bibnamefont {Chen}}, \ and\ \bibinfo {author}
  {\bibfnamefont {Jian-Wei}\ \bibnamefont {Pan}},\ }\bibfield  {title}
  {\enquote {\bibinfo {title} {Realization of an ideal weyl semimetal band in a
  quantum gas with 3d spin-orbit coupling},}\ }\href {\doibase
  10.1126/science.abc0105} {\bibfield  {journal} {\bibinfo  {journal}
  {Science}\ }\textbf {\bibinfo {volume} {372}},\ \bibinfo {pages} {271--276}
  (\bibinfo {year} {2021})}\BibitemShut {NoStop}%
\bibitem [{\citenamefont {Edmonds}(1957)}]{Edmonds1957}%
  \BibitemOpen
  \bibfield  {author} {\bibinfo {author} {\bibfnamefont {Alan~Robert}\
  \bibnamefont {Edmonds}},\ }\href@noop {} {\emph {\bibinfo {title} {Angular
  momentum in quantum mechanics}}}\ (\bibinfo  {publisher} {Princeton
  university press},\ \bibinfo {year} {1957})\BibitemShut {NoStop}%
\bibitem [{\citenamefont {Jain}(1989)}]{Jain1989}%
  \BibitemOpen
  \bibfield  {author} {\bibinfo {author} {\bibfnamefont {J.~K.}\ \bibnamefont
  {Jain}},\ }\bibfield  {title} {\enquote {\bibinfo {title} {Composite-fermion
  approach for the fractional quantum hall effect},}\ }\href {\doibase
  10.1103/PhysRevLett.63.199} {\bibfield  {journal} {\bibinfo  {journal} {Phys.
  Rev. Lett.}\ }\textbf {\bibinfo {volume} {63}},\ \bibinfo {pages} {199--202}
  (\bibinfo {year} {1989})}\BibitemShut {NoStop}%
\end{thebibliography}

\begin{thebibliography}{6}%
\makeatletter
\providecommand \@ifxundefined [1]{%
 \@ifx{#1\undefined}
}%
\providecommand \@ifnum [1]{%
 \ifnum #1\expandafter \@firstoftwo
 \else \expandafter \@secondoftwo
 \fi
}%
\providecommand \@ifx [1]{%
 \ifx #1\expandafter \@firstoftwo
 \else \expandafter \@secondoftwo
 \fi
}%
\providecommand \natexlab [1]{#1}%
\providecommand \enquote  [1]{``#1''}%
\providecommand \bibnamefont  [1]{#1}%
\providecommand \bibfnamefont [1]{#1}%
\providecommand \citenamefont [1]{#1}%
\providecommand \href@noop [0]{\@secondoftwo}%
\providecommand \href [0]{\begingroup \@sanitize@url \@href}%
\providecommand \@href[1]{\@@startlink{#1}\@@href}%
\providecommand \@@href[1]{\endgroup#1\@@endlink}%
\providecommand \@sanitize@url [0]{\catcode `\\12\catcode `\$12\catcode
  `\&12\catcode `\#12\catcode `\^12\catcode `\_12\catcode `\%12\relax}%
\providecommand \@@startlink[1]{}%
\providecommand \@@endlink[0]{}%
\providecommand \url  [0]{\begingroup\@sanitize@url \@url }%
\providecommand \@url [1]{\endgroup\@href {#1}{\urlprefix }}%
\providecommand \urlprefix  [0]{URL }%
\providecommand \Eprint [0]{\href }%
\providecommand \doibase [0]{http://dx.doi.org/}%
\providecommand \selectlanguage [0]{\@gobble}%
\providecommand \bibinfo  [0]{\@secondoftwo}%
\providecommand \bibfield  [0]{\@secondoftwo}%
\providecommand \translation [1]{[#1]}%
\providecommand \BibitemOpen [0]{}%
\providecommand \bibitemStop [0]{}%
\providecommand \bibitemNoStop [0]{.\EOS\space}%
\providecommand \EOS [0]{\spacefactor3000\relax}%
\providecommand \BibitemShut  [1]{\csname bibitem#1\endcsname}%
\let\auto@bib@innerbib\@empty
\bibitem [{\citenamefont {Edmonds}(1957)}]{Edmonds1957SM}%
  \BibitemOpen
  \bibfield  {author} {\bibinfo {author} {\bibfnamefont {Alan~Robert}\
  \bibnamefont {Edmonds}},\ }\href@noop {} {\emph {\bibinfo {title} {Angular
  momentum in quantum mechanics}}}\ (\bibinfo  {publisher} {Princeton
  university press},\ \bibinfo {year} {1957})\BibitemShut {NoStop}%
\bibitem [{\citenamefont {Wu}\ and\ \citenamefont {Yang}(1977)}]{Wu1977SM}%
  \BibitemOpen
  \bibfield  {author} {\bibinfo {author} {\bibfnamefont {Tai~Tsun}\
  \bibnamefont {Wu}}\ and\ \bibinfo {author} {\bibfnamefont {Chen~Ning}\
  \bibnamefont {Yang}},\ }\bibfield  {title} {\enquote {\bibinfo {title} {Some
  properties of monopole harmonics},}\ }\href {\doibase
  10.1103/PhysRevD.16.1018} {\bibfield  {journal} {\bibinfo  {journal} {Phys.
  Rev. D}\ }\textbf {\bibinfo {volume} {16}},\ \bibinfo {pages} {1018--1021}
  (\bibinfo {year} {1977})}\BibitemShut {NoStop}%
\bibitem [{\citenamefont {Stone}\ and\ \citenamefont
  {Goldbart}(2009)}]{Stone2009SM}%
  \BibitemOpen
  \bibfield  {author} {\bibinfo {author} {\bibfnamefont {Michael}\ \bibnamefont
  {Stone}}\ and\ \bibinfo {author} {\bibfnamefont {Paul}\ \bibnamefont
  {Goldbart}},\ }\href {\doibase 10.1017/cbo9780511627040} {\emph {\bibinfo
  {title} {Mathematics for Physics}}}\ (\bibinfo  {publisher} {Cambridge
  University Press},\ \bibinfo {year} {2009})\BibitemShut {NoStop}%
\bibitem [{\citenamefont {Wilczek}\ and\ \citenamefont
  {Zee}(1983)}]{Wilczek1983SM}%
  \BibitemOpen
  \bibfield  {author} {\bibinfo {author} {\bibfnamefont {Frank}\ \bibnamefont
  {Wilczek}}\ and\ \bibinfo {author} {\bibfnamefont {A.}~\bibnamefont {Zee}},\
  }\bibfield  {title} {\enquote {\bibinfo {title} {Linking numbers, spin, and
  statistics of solitons},}\ }\href {\doibase 10.1103/PhysRevLett.51.2250}
  {\bibfield  {journal} {\bibinfo  {journal} {Phys. Rev. Lett.}\ }\textbf
  {\bibinfo {volume} {51}},\ \bibinfo {pages} {2250--2252} (\bibinfo {year}
  {1983})}\BibitemShut {NoStop}%
\bibitem [{\citenamefont {Mermin}\ and\ \citenamefont {Ho}(1976)}]{Mermin1976SM}%
  \BibitemOpen
  \bibfield  {author} {\bibinfo {author} {\bibfnamefont {N.~D.}\ \bibnamefont
  {Mermin}}\ and\ \bibinfo {author} {\bibfnamefont {Tin-Lun}\ \bibnamefont
  {Ho}},\ }\bibfield  {title} {\enquote {\bibinfo {title} {Circulation and
  angular momentum in the {$A$} phase of superfluid helium-3},}\ }\href
  {\doibase 10.1103/PhysRevLett.36.594} {\bibfield  {journal} {\bibinfo
  {journal} {Phys. Rev. Lett.}\ }\textbf {\bibinfo {volume} {36}},\ \bibinfo
  {pages} {594--597} (\bibinfo {year} {1976})}\BibitemShut {NoStop}%
\bibitem [{\citenamefont {Kamien}(2002)}]{Kamien2002SM}%
  \BibitemOpen
  \bibfield  {author} {\bibinfo {author} {\bibfnamefont {Randall~D.}\
  \bibnamefont {Kamien}},\ }\bibfield  {title} {\enquote {\bibinfo {title} {The
  geometry of soft materials: a primer},}\ }\href {\doibase
  10.1103/revmodphys.74.953} {\bibfield  {journal} {\bibinfo  {journal}
  {Reviews of Modern Physics}\ }\textbf {\bibinfo {volume} {74}},\ \bibinfo
  {pages} {953--971} (\bibinfo {year} {2002})}\BibitemShut {NoStop}%
\end{thebibliography}
%

\end{document}